\documentclass[10pt,twocolumn,letterpaper]{article}

\usepackage{cvpr}              
\definecolor{cvprblue}{rgb}{0.21,0.49,0.74}
\usepackage[pagebackref,breaklinks,colorlinks,allcolors=cvprblue]{hyperref}
\usepackage{algorithm}          %
\usepackage{algpseudocode}      %
\usepackage{multirow}
\definecolor{lavender}{RGB}{230,228,248}
\usepackage{tikz}
\usepackage{booktabs}
\usepackage{multirow}
\usepackage{graphicx}
\usepackage{xcolor}
\def\paperID{} 
\def\confName{CVPR}
\def\confYear{2026}

\title{Keep the Core: Adversarial Priors for Significance-Preserving \\ Brain MRI Segmentation}

\author{Feifei Zhang\\
Institution1\\
Institution1 address\\
{\tt\small firstauthor@i1.org}
\and
Second Author\\
Institution2\\
First line of institution2 address\\
{\tt\small secondauthor@i2.org}
}

\author{Feifei Zhang$^{1, 2, 3}$, Zhenhong Jia$^{1, 2, *}$, Sensen Song$^{1, 2}$, Fei Shi$^{1, 2, 3}$, Aoxue Chen$^{5}$, Dayong Ren$^{4,}$ \thanks{Corresponding author.}  \\\small
	$^{1}$School of Computer Science and Technology, Xinjiang University, Urumqi, 830046, China, \\ \small
	$^{2}$ Key Laboratory of Signal Detection and Processing, Xinjiang University, Urumqi, 830046, China,\\\small
	$^{3}$ Xinjiang Multimodal Intelligent Processing and Information Security Engineering Technology \\ \small Research Center, Urumqi, 830046, China,\\ \small
	$^{4}$ National Key Laboratory for Novel Software Technology, Nanjing University, Nanjing 210023, China, \\ \small
	$^{5}$ Department of Neurology, Jiangsu Province Hospital, \\ \small The First Affiliated Hospital of Nanjing Medical University, Nanjing 210029, China.
}

\begin{document}
\maketitle
\begin{abstract}
Medical image segmentation is constrained by sparse pathological annotations. Existing augmentation strategies, from conventional transforms to random masking for self-supervision, are feature-agnostic: they often corrupt critical diagnostic semantics or fail to prioritize essential features. We introduce "Keep the Core," a novel data-centric paradigm that uses adversarial priors to guide both augmentation and masking in a significance-preserving manner. Our approach uses SAGE (Sparse Adversarial Gated Estimator), an offline module identifying minimal tokens whose micro-perturbation flips segmentation boundaries. SAGE forges the Token Importance Map $W$ by solving an adversarial optimization problem to maximally degrade performance, while an $\ell_1$ sparsity penalty encourages a compact set of sensitive tokens. The online KEEP (Key-region Enhancement \& Preservation) module uses $W$ for a two-pronged augmentation strategy: (1) Semantic-Preserving Augmentation: High-importance tokens are augmented, but their original pixel values are strictly restored. (2) Guided-Masking Augmentation: Low-importance tokens are selectively masked for an $\text{MAE}$-style reconstruction, forcing the model to learn robust representations from preserved critical features. "Keep the Core" is backbone-agnostic with no inference overhead. Extensive experiments show SAGE's structured priors and KEEP's region-selective mechanism are highly complementary, achieving state-of-the-art segmentation robustness and generalization on 2D medical datasets.
\vspace{-0.8cm}
\end{abstract}    
\section{Introduction}
\label{sec:intro}
Medical image segmentation serves as a cornerstone for modern clinical diagnosis, prognostic assessment, and surgical planning \cite{SAM,12,15}, providing quantitative morphological and spatial information from CT and MRI for tumor delineation and organ mapping. However, the remarkable performance of these models relies heavily on large-scale, high-quality, pixel-level annotated datasets. In the medical domain, acquiring such data faces two major bottlenecks: (1) Data Scarcity: Privacy regulations, equipment costs, and data silos make acquiring large-scale medical images challenging. (2) Annotation Scarcity: Labeling medical images requires expert radiologists or pathologists, a process that is both time-consuming and expensive. More critically, pathological annotations are often extremely sparse spatially—for instance, micro-nodules, early-stage polyps, or diffuse lesions may occupy less than 1\% of an entire scan \cite{hesamian2019deep,13}. To mitigate data scarcity, Data Augmentation (DA) has become a standard component in deep learning training \cite{shorten2019survey}. Traditional DA methods, such as affine transformations (rotation, scaling) and color jittering (brightness, contrast), are vital for improving model robustness to variations in position and illumination.

However, when more advanced regional augmentation methods (e.g., Cutout \cite{devries2017cutout}, Random Erasing \cite{zhong2020random}, or even Mixup \cite{zhang2018mixup}) are blindly applied to medical images, they often yield catastrophic results. These methods, designed for natural images, assume high information redundancy. In medical images, when a random occlusion patch happens to cover a tiny, critical lesion, the sample's critical diagnostic semantics are entirely destroyed \cite{chlap2021review}. The model learns from this "corrupted" sample may be noise, artifacts, or an incorrect "background prior." This issue of semantics-agnostic data manipulation is not limited to traditional DA; it extends directly to modern self-supervised strategies like Masked Image Modeling.

As another solution to annotation scarcity, Self-Supervised Learning (SSL), particularly Masked Image Modeling (MIM) \cite{bao2021beit} as exemplified by the Masked Autoencoder (MAE) \cite{he2022mae} has shown immense potential. Yet, its core augmentation technique—random masking—suffers from the same 'blind spot' in medical imaging. In natural images, information is highly redundant (e.g., sky, grass), and random masking forces the model to learn global context. In medical images, a single 16x16 token might contain the entire early-stage lesion. MAE's random policy has a high probability of masking all critical pathological tokens. This leads to a paradox: the model spends significant pre-training time learning to perfectly reconstruct healthy anatomical structures (i.e., the "background") but learns little about the critical lesion representations \cite{tang2022self}. This strategy is not only inefficient but also creates a fundamental bias between the learned representation and the downstream segmentation task (i.e., localizing the lesion).

These challenges point to a common root: the field lacks a unified, semantics-aware augmentation paradigm. Both conventional transforms and SSL masking operate 'blindly,' failing to distinguish 'critical' from 'non-critical' tokens. We propose a more fundamental and robust definition of "importance." We argue that: A token's "importance" is not defined by its visual saliency, but by its "Adversarial Vulnerability" to the model's decision boundary. In other words, if applying a tiny, imperceptible perturbation to a token can cause the model's segmentation output (e.g., Dice score) to catastrophically collapse (i.e., "flip the segmentation boundary"), then that token carries the most critical diagnostic information. In medical images, these "vulnerable" tokens correspond precisely to the subtle textures, ambiguous borders, or micro-structures that the model relies on to distinguish "pathology" from "health." These regions are the true cornerstones of the model's decision-making and are the most easily destroyed by DA or ignored by MAE. Based on this core insight, we propose "Keep the Core," a novel data-centric paradigm that leverages adversarial priors to intelligently guide both data augmentation and masked self-supervised learning.

Our approach begins with the SAGE (Sparse Adversarial Gated Estimator), an offline module whose sole objective is to compute this "adversarial vulnerability." SAGE operates on a token grid, and its core task is to solve a sparse adversarial optimization problem: to find the minimum set of tokens where applying a micro-perturbation is sufficient to "flip" the model's segmentation boundary. The output of SAGE is a normalized Token Importance Map $W$, which precisely quantifies each token's contribution to the model's decision, thereby effectively localizing the most sensitive pathological regions.

In the online training stage, the KEEP (Key-region Enhancement \& Preservation) module utilizes the offline-generated $W$ map to execute an adaptive, two-pronged intervention on the standard training pipeline (including DA and MAE):
1. Preservation: "Augment-and-Restore" for Critical Tokens For high-importance tokens identified by $W$ (i.e., the "core" lesion areas): we first apply standard intensity and affine transformations (DA) to the entire image. Then, we execute a critical "restore" operation to ensure the integrity and high-fidelity of core semantic features, while still allowing the model to observe the core features in varied augmented contexts. 2. Guided Masking for Non-Critical Tokens. For low-importance tokens in $W$ (e.g., background, healthy tissue): we perform guided MAE-style masking. We prioritize sampling tokens for masking from this "non-critical" pool. This forces the model to solve a more valuable task: "reconstruct the surrounding non-critical context, given the (visible) core pathological features." This drastically improves representation learning efficiency and forces the model to learn robust representations that rely on the preserved critical features. The "Keep the Core" framework is backbone-agnostic and introduces no inference overhead. We conduct extensive experiments on multiple 2D medical segmentation datasets (e.g., \cite{menze2015brats, heller2021state}). Our contributions are summarized as follows:
\begin{itemize}
    \item We introduce adversarial vulnerability as a new definition of feature importance, which is more aligned with the segmentation boundary than visual saliency.
    \item We propose a two-stage SAGE-KEEP framework where an offline-generated adversarial map ($W$) guides an online dual-path strategy of semantic preservation and guided masking.
\end{itemize}

\section{Related Work}
\label{sec:related}

Our work is positioned at the intersection of network architecture and data-centric learning strategies. We uniquely categorize both traditional Data Augmentation and the masking policies within Self-Supervised Learning as two facets of a broader augmentation paradigm, which our work aims to unify.

\begin{figure*}[t!]
\centering
\includegraphics[scale=0.65, trim=4.5cm 10cm 5.5cm 2cm, clip]{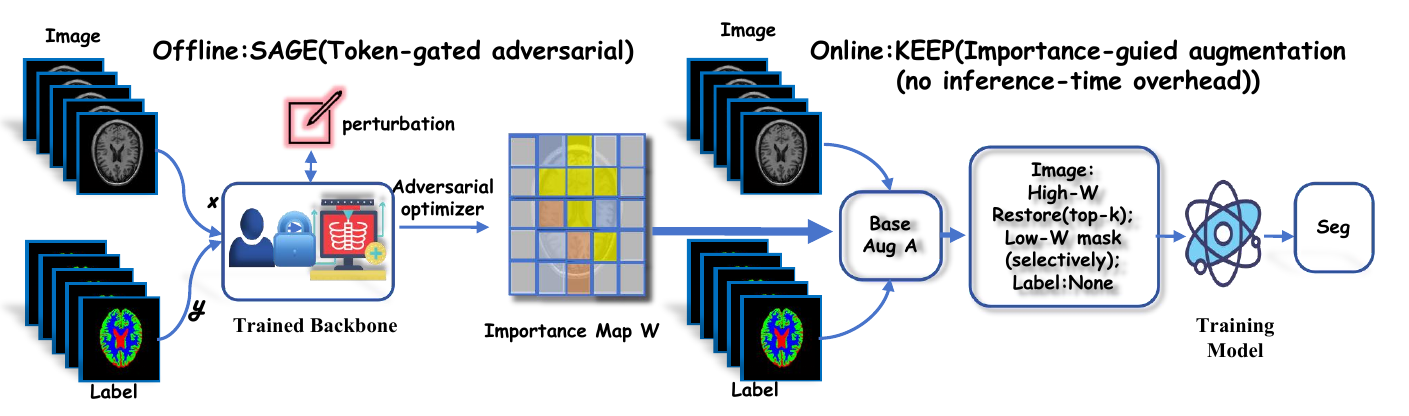}
\caption{"Keep the Core". \textbf{(Left)} The offline SAGE (Sparse Adversarial Gating Evaluator) solves an adversarial optimization problem to generate a Token Importance Map ($W$). \textbf{(Right)} The online KEEP (Key-region Enhancement and Preservation) module leverages $W$ to apply its two-pronged strategy: (1) "Augment-and-Restore" for critical tokens and (2) "Guided Masking" for non-critical tokens.}
\label{fig1}
\end{figure*}

\subsection{Architectural Evolution}
With the rapid progress of deep learning, a significant body of research has focused on improving segmentation performance through architectural innovation. Early CNN-based models such as FCN \cite{fcn}, U-Net \cite{ronneberger2015unet} and its variants (U-Net++ \cite{unet++}, U-Net3+ \cite{unet3+}, nnU-Net \cite{nnunet}) leveraged encoder–decoder architectures with multi-level skip connections to improve multi-scale feature fusion and lesion localization \cite{can}, but remained limited by local receptive fields and insufficient global context modeling \cite{17visionmamba, 1, 2, 3, 4}. 
Transformer-based approaches \cite{15ViT}, including TransUNet \cite{chen2021transunet}, UTNet \cite{utnet}, ViT-UNet \cite{vit-unet}, and Swin-UNet \cite{swin-unet}, introduced self-attention to aggregate global information, yet their quadratic complexity in self-attention incurs substantial computational costs \cite{dualmambanet, 5, 6, 7, 8, 9, 10, 11}. 
More recently, state space models (SSMs) \cite{16mamba} such as Mamba have emerged as a promising alternative, offering linear complexity while maintaining strong long-range dependency modeling. Vision Mamba backbones and U-Net variants \cite{liu2024swin} as well as hybrid designs like LMa-UNet \cite{lma-unet}, LocalMamba \cite{localmamba}, and SegMamba \cite{Segmamba} further enhance spatial awareness and multi-scale fusion.
While architectural advancements are crucial, our work is complementary and backbone-agnostic. We address a more fundamental challenge: how models learn from limited and sparsely annotated data, regardless of the specific architecture employed.

\subsection{DA and Semantics Preservation}
Data Augmentation (DA) is a standard technique to combat data scarcity \cite{shorten2019survey}. Simple affine and color transformations are beneficial. However, regional dropout methods, such as Cutout \cite{devries2017cutout} and Random Erasing \cite{zhong2020random}, designed for information-redundant natural images, often fail in medical contexts. As noted by \cite{chlap2021review}, randomly occluding a tiny, critical lesion destroys the sample's diagnostic semantics, forcing the model to learn from corrupted labels.
Several works have attempted "semantics-aware" augmentation. For instance, \cite{gong2021keepaugment} used saliency maps (e.g., \cite{simonyan2014deep, selvaraju2017grad}) to prevent augmentation in important areas. However, extensive research \cite{adebayo2018sanity} shows that saliency maps are unreliable. At best, they reflect only \textit{where} a model is looking, not what is \textit{critical} to its decision boundary \cite{kindermans2019reliability}. Our approach differs by defining importance not via post-hoc attribution, but directly through adversarial vulnerability, which is inherently tied to the decision boundary.

\subsection{Masked Image Modeling in Medical Imaging}
Masked Image Modeling (MIM), exemplified by the Masked Autoencoder (MAE) \cite{he2022mae}, has become a powerful self-supervised learning (SSL) strategy. We analyze it here as a modern form of data augmentation, where the core augmentation technique is the "random masking" policy. However, this "random masking" is suboptimal for medical images. Pathological features (e.g., micro-nodules) are often small and sparse. A 75\% random masking has a high probability of masking all critical features, forcing the model to spend its capacity learning to reconstruct healthy, "background" anatomy \cite{tang2022self}. This creates a bias, where the pre-trained representation is proficient at anatomy but naive about pathology. Some works have explored guided masking, but often rely on saliency or visual features. Our SAGE module provides a more robust, non-random prior, guiding the MAE task to focus on reconstructing the \textit{context} around the critical features we explicitly preserve.

\begin{figure}[!ht]
    \centering
    \includegraphics[width=0.95\linewidth, trim=3.5cm 11cm 12cm 1cm, clip]{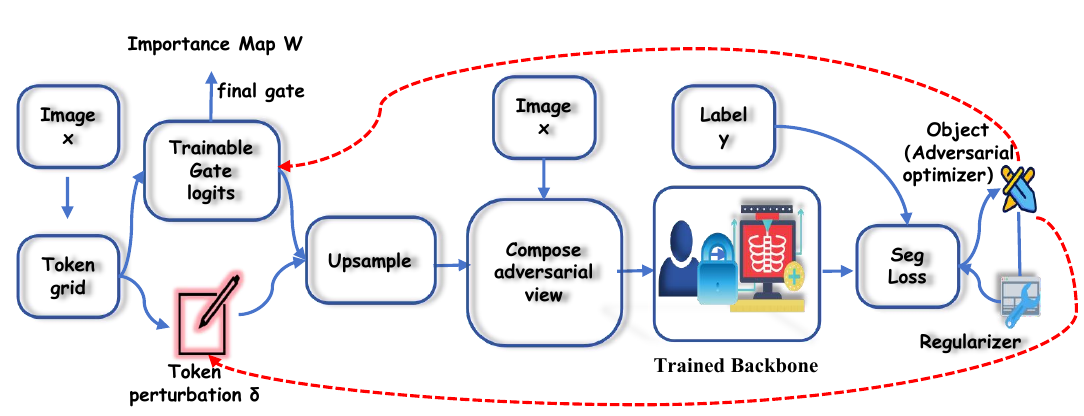}
    \caption{\textbf{The Offline SAGE Module.} SAGE freezes the segmentation network and optimizes a learnable Gated Mask $m$ and a Perturbation $\delta$. By minimizing the combined adversarial and sparsity objective, it forces the mask to converge on the most sensitive pathological regions, producing the Token Importance Map $W$.}
    \label{fig2}
\end{figure}
\section{Method}
\label{sec:meth}

\subsection{Overview}
We introduce \textbf{Keep the Core}, a novel framework designed to reconcile the conflict between data scarcity and semantic integrity in medical image segmentation. As illustrated in \cref{fig1}, our framework decouples the process into two distinct yet complementary phases:
(1) \textbf{Offline Vulnerability Assessment (SAGE)}: Before the main training loop, we perform a computationally intensive "stress test" on the training data. By solving a sparse adversarial optimization problem, the SAGE module identifies the minimal set of image tokens—termed the "Core"—that are most critical for the model's decision boundary.
(2) \textbf{Online Robustness Enhancement (KEEP)}: During the efficient online training phase, the KEEP module utilizes these pre-computed importance priors to execute a meta-augmentation paradigm. It does not replace standard augmentations, but intelligently guides them by selectively protecting the fragile core tokens while aggressively challenging the non-critical context, achieving a balance between feature preservation and robustness without inference-time overhead.

\subsection{SAGE: Sparse Adversarial Gating Evaluator}
The primary objective of SAGE is to generate a pixel-level \emph{Token Importance Map} $W$ for each training sample $(x, y)$. Unlike heuristic saliency methods (e.g., Grad-CAM) which only reflect where the model "looks," SAGE actively "probes" the model to find where it is "vulnerable." We hypothesize that the most critical diagnostic features are those whose perturbation leads to the most significant collapse in segmentation performance.

\subsubsection{Adversarial Gating Formulation}
As depicted in \cref{fig2}, SAGE operates on a frozen, pre-trained segmentation oracle $f(\cdot)$. Given an input image $x \in \mathbb{R}^{H \times W}$, we discretize the spatial domain into a grid of tokens with resolution $(H_t, W_t)$. To locate critical tokens, we introduce two learnable tensor components:
\begin{itemize}
    \item \textbf{Latent Gate $G \in \mathbb{R}^{H_t \times W_t}$}: A real-valued parameter map that controls the spatial selection of tokens.
    \item \textbf{Perturbation $\delta \in \mathbb{R}^{C \times H_t \times W_t}$}: A learnable noise tensor bounded by an $\ell_\infty$ norm constraint $\|\delta\|_\infty \le \epsilon$.
\end{itemize}
To enable gradient-based optimization of the discrete selection process, we employ a differentiable relaxation using temperature annealing. The binary-like mask $m$ is computed as $m = \sigma(G / T)$, where $\sigma(\cdot)$ is the sigmoid function and $T$ is a temperature scalar that decays over time. The resulting adversarial sample $x_{\text{adv}}$ is synthesized by injecting the perturbation only into the gated tokens:
\begin{equation}
\label{eq:adv_sample}
x_{\text{adv}} = \mathrm{clamp}\left(x + \mathcal{U}(m) \odot \mathcal{U}(\delta)\right),
\end{equation}
where $\mathcal{U}(\cdot)$ denotes nearest-neighbor upsampling to match the image resolution, and $\odot$ represents the Hadamard product.

\subsubsection{Sparse Optimization Objective}
SAGE formulates importance discovery as a \emph{Min-Max} optimization game. We seek to maximize the segmentation error (attack) while simultaneously minimizing the area of the selected tokens (sparsity). This ensures that $W$ captures only the \emph{minimal sufficient} features required to flip the prediction. The total objective $\mathcal{L}_{\text{SAGE}}$ is defined as:
\begin{align}
    \mathcal{L}_{\text{attack}} &= \lambda_{\text{ce}}\mathcal{L}_{\text{CE}}(f(x_{\text{adv}}), y) + \lambda_{\text{dice}}\mathcal{L}_{\text{Dice}}(f(x_{\text{adv}}), y), \\
    \mathcal{L}_{\text{sparse}} &= \mu_{\text{sparse}} \|m\|_1 + \beta_{\delta} \|\delta\|_1, \\
    \min_{G, \delta} \mathcal{L}_{\text{SAGE}} &= \min_{G, \delta} \left( - \mathcal{L}_{\text{attack}} + \mathcal{L}_{\text{sparse}} \right).
\end{align}
Here, the $\ell_1$ penalty on $m$ is crucial: it forces the gate to "shut down" for all pixels except those strictly necessary to degrade the model's performance. As shown in \cref{alg:sage}, we iteratively update $G$ and $\delta$. Upon convergence, the final Importance Map is derived as $W = \sigma(G / T_{\text{end}})$, effectively highlighting the "Achilles' heel" of the input sample.

\begin{algorithm}[t]
\caption{SAGE: Offline Importance Generation}
\label{alg:sage}
\small
\textbf{Input:} Image $x$, Label $y$, Frozen Oracle $f(\cdot)$. \\
\textbf{Params:} Budget $\epsilon$, Temp $T_{start} \to T_{end}$, Steps $N$. \\
\textbf{Output:} Token Importance Map $W$.
\begin{algorithmic}[1]
\State Initialize $G \leftarrow \mathbf{0}$, $\delta \leftarrow \mathbf{0}$.
\For{$step = 1$ \textbf{to} $N$}
    \State $T \leftarrow \text{Anneal}(T_{start}, T_{end}, step)$
    \State $m \leftarrow \sigma(G / T)$ \Comment{Generate soft mask}
    \State $x_{\text{adv}} \leftarrow \text{clamp}(x + \mathcal{U}(m) \odot \mathcal{U}(\delta))$
    \State $\mathcal{L} \leftarrow - \mathcal{L}_{seg}(f(x_{\text{adv}}), y) + \mu \|m\|_1 + \beta \|\delta\|_1$
    \State Update $G, \delta$ via Adam to minimize $\mathcal{L}$
    \State $\delta \leftarrow \text{clip}(\delta, -\epsilon, \epsilon)$ \Comment{Enforce budget}
\EndFor
\State \Return $W \leftarrow \sigma(G / T_{end})$
\end{algorithmic}
\end{algorithm}

\begin{algorithm}[t]
\caption{KEEP: Online Guided Augmentation}
\label{alg:keep}
\small
\textbf{Input:} Image $x$, Importance Map $W$, Standard Augmentor $\mathcal{A}$.
\textbf{Params:} Top-K value $K$, Optional mask threshold $\tau_{low}$.
\textbf{Output:} Final augmented image $x_{final}$.
\begin{algorithmic}[1]
\State $x_{aug} \leftarrow \mathcal{A}(x)$ \Comment{Apply standard DA (e.g., affine, color)}
\State $S_{local} \leftarrow \text{Pool}(W)$ \Comment{Compute token-level importance scores}
\State
\State \textcolor{gray}{// Path 1: Preserve Critical Core (Fig. \ref{fig3} Top)}
\State $M_{core} \leftarrow \text{TopK\_Mask}(S_{local}, K)$ \Comment{Get mask for Top-K tokens}
\State $x' \leftarrow x_{aug} \odot (1 - M_{core}) + x \odot M_{core}$ \Comment{Restore original pixels}
\State
\State \textcolor{gray}{// Path 2: Optional Guided Masking (Fig. \ref{fig3} Bottom)}
\State $M_{mask} \leftarrow \mathbf{0}$
\If {$\tau_{low} > 0$} \Comment{Masking is optional}
    \State $M_{potential} \leftarrow \mathbb{I}(S_{local} < \tau_{low})$ \Comment{Find low-score tokens}
    \State $M_{mask} \leftarrow \text{SampleSubset}(M_{potential})$ \Comment{Sample from non-critical pool}
\EndIf
\State $x_{final} \leftarrow x' \odot (1 - M_{mask})$ \Comment{Apply mask (or identity if $M_{mask}$ is 0)}
\State \Return $x_{final}$
\end{algorithmic}
\end{algorithm}

\subsection{KEEP: Key-region Enhancement and Preservation}
The KEEP module operationalizes our unified augmentation paradigm. It functions as an intelligent wrapper around the standard data augmentation pipeline $\mathcal{A}(\cdot)$. As illustrated in \cref{fig3} and formalized in \cref{alg:keep}, it follows a "Protect-and-Challenge" philosophy: safeguarding high-value pathological semantics while forcing the model to infer context from incomplete data.

After a standard stochastic augmentation $\mathcal{A}(\cdot)$ (e.g., random intensity shifts, blurring) is applied to produce $x_{aug}$, KEEP applies the following two interventions:

\subsubsection{Core Preservation (Augment-and-Restore)}
Standard augmentations can inadvertently destroy subtle pathological details (e.g., washing out the texture of a micro-nodule). To prevent this, we identify the set of Top-K tokens $M_{core}$ with the highest importance scores (derived from $W$). As shown in the upper branch of \cref{fig3}, we strictly \emph{restore} the original pixel values from the pre-augmentation image $x$ within these tokens:
\begin{equation}
    x' = x_{aug} \odot (1 - M_{core}) + x \odot M_{core}.
\end{equation}
This "Augment-and-Restore" strategy ensures that the model always sees high-fidelity data for the most diagnostically relevant features, while the surrounding context remains augmented to provide variability.

\subsubsection{Optional Guided Context Masking}
\begin{figure}[t!]
    \centering
    \includegraphics[width=0.95\linewidth, trim=1cm 8cm 20cm 1.5cm, clip]{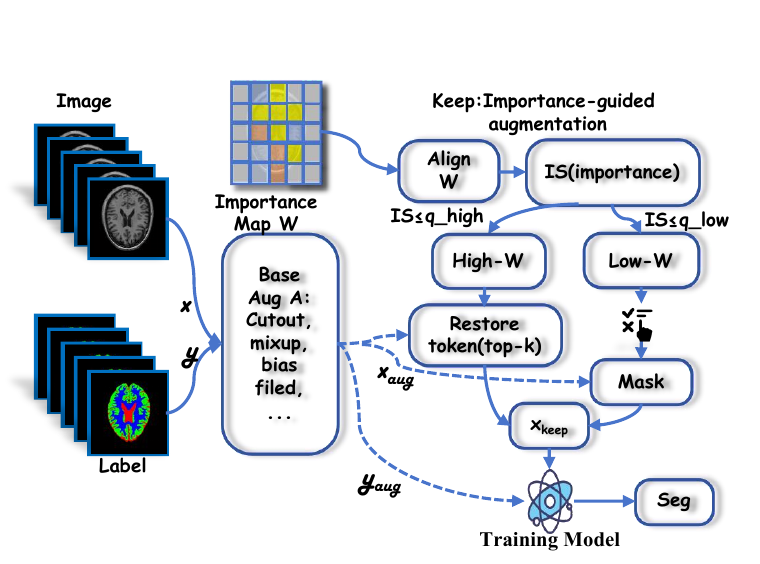}
    \caption{\textbf{The Online KEEP Module.} Using the Importance Map $W$, KEEP performs a two-step intervention after standard augmentation. \textbf{Path 1 (Top):} Critical tokens (Red) are restored to their original values to prevent semantic corruption. \textbf{Path 2 (Bottom):} Non-critical background tokens (Blue) are optionally masked to encourage robust global context learning.}
    \label{fig3}
\end{figure}

Conversely, medical background tokens are often highly redundant. To prevent the model from relying on trivial correlations and to enhance representation learning, we optionally employ a guided masking strategy. We identify a pool of non-critical tokens $M_{potential}$ (e.g., those with importance scores below a threshold $\tau_{low}$). As shown in the lower branch of \cref{fig3}, we randomly sample a subset $M_{mask}$ from this pool and mask them (e.g., set to zero or a fill value):
\begin{equation}
    x_{final} = x' \odot (1 - M_{mask}).
\end{equation}
This effectively constructs a hard reconstruction task, compelling the network to learn robust global representations by inferring the missing context solely from the preserved core features. If this optional step is skipped, $x_{final} = x'$.

\section{Experiments}
\label{sec:expe}

\subsection{Experimental Setup}
\label{sec:setup}

\paragraph{Datasets.}
We validate our framework on two public benchmarks covering  2D visual modality:
\begin{itemize}
    \item \textbf{OASIS-1 \cite{oasis1}:} Derived from the Open Access Series of Imaging Studies, this dataset comprises T1-weighted MRI scans from 421 subjects (aged 18--96). The images were acquired with a resolution of $176 \times 208$ pixels and a slice thickness of 1.25 mm (TR=$9.7$ ms, TE=$4.0$ ms, TI=$20$ ms). Following standard protocols, we utilize the provided manual segmentation masks for Cerebrospinal Fluid (CSF), Grey Matter (GM), and White Matter (WM).
    \item \textbf{MRBrainS13 \cite{mrbrains}:} From the MICCAI 2013 challenge, this dataset contains multi-sequence scans of 20 subjects acquired on a 3.0T Philips Achieva scanner. It provides T1 (TR: $7.9$ ms, TE: $4.5$ ms), T1-IR, and T2-FLAIR sequences. All scans are co-registered and bias-corrected with a voxel spacing of $0.96 \times 0.96 \times 3.00$ mm. We utilize the provided labels for CSF, GM, and WM segmentation.
\end{itemize}
\vspace{-0.5cm}

\paragraph{Implementation Details.}
Our framework is implemented in PyTorch and trained on a single NVIDIA A100 GPU (40GB). We employ the AdamW optimizer with an initial learning rate of $1 \times 10^{-3}$ and a cosine annealing decay schedule. The batch size is set to 16. Models are trained for 50 epochs on OASIS-1 and 200 epochs on MRBrainS13 to ensure convergence. For evaluation, we employ an early-stopping strategy where the checkpoint achieving the highest Dice score on the validation set is selected. 
\textbf{Hyperparameters:} Following \cite{devries2017improved, yang2025ipf}, the annealing parameter $\alpha$ in SAGE is linearly increased from $\alpha_{init}=0.1$ to $\alpha_{end}=10$ to encourage binary mask convergence. The temperature parameter $\tau$ is set to 0.6 across all experiments. Crucially, these hyperparameters are fixed across different backbones, demonstrating the robust and plug-and-play nature of our framework.
\vspace{-0.5cm}

\paragraph{Baseline Methods.}
To ensure a rigorous comparison, we benchmark \textit{Keep the Core} against a comprehensive suite of augmentation strategies across three representative backbones: UNet, SwinUNet \cite{swin-unet}, and SwinUMamba \cite{liu2024swin}. The comparison methods are categorized into:
(1) \textbf{Pixel-level Augmentation:} Gaussian noise \cite{tellez2019quantifying}, Gaussian blur \cite{hussain2018differential}, Gamma correction \cite{sun2021robust}, Brightness/Contrast adjustment \cite{sirazitdinov2019data}, and Bias field perturbation \cite{chen2020realistic}.
(2) \textbf{Token-level Augmentation:} Random Erasing \cite{zhong2020random}, Cutout \cite{devries2017cutout}, Mixup \cite{zhang2018mixup}, Cutmix \cite{yun2019cutmix}, HSMix \cite{sun2025hsmix}, MRS \cite{huang2025rethinking}, and ABD \cite{chi2024adaptive}.
All baselines are re-implemented using their officially reported optimal configurations. To guarantee a fair and rigorous comparison, all models (baselines and ours) are trained from scratch using an identical optimization protocol, data splits, and hyperparameter settings, thereby isolating the performance contribution of the augmentation strategy. 

\begin{table*}[t!]
\centering
\caption{Comprehensive evaluation on OASIS-1 dataset. 'Keep the Core' (KC) methods are compared to their non-KC counterparts(e.g., KC-Gaussian vs. Gaussian Noise). Metric improvements are marked with a top-right red arrow (\textcolor{red}{$\uparrow$}), degradations with a bottom-right green arrow (\textcolor{green}{$\downarrow$}).}
\resizebox{\textwidth}{!}{%
\renewcommand{\arraystretch}{0.7}
\begin{tabular}{cccccccccccccc}
\cmidrule[1.5pt]{1-14}
\multirow{2}{*}{\textbf{Backbone}} & \multirow{2}{*}{\textbf{Augmentation}} & \multicolumn{3}{c}{\textbf{Dice} $\uparrow$} & \multicolumn{3}{c}{\textbf{HD95(mm)}$\downarrow$} & \multicolumn{3}{c}{\textbf{ASD(mm)}$\downarrow$} & \multicolumn{3}{c}{\textbf{IOU}$\uparrow$} \\
\cmidrule[1pt]{3-14}
& & CSF & GM & WM & CSF & GM & WM & CSF & GM & WM & CSF & GM & WM \\
\cmidrule[1pt]{1-14}

\multirow{5}{*}{\textbf{UNet}} 
& Baseline & 0.9062 & 0.9134 & 0.9181 & 1.1267 & 1.2138 & 2.0824 & 0.2581 & 0.3611 & 0.6198 & 0.8442 & 0.8599 & 0.8688 \\
& Gaussian Noise & 0.9038 & 0.9141 & 0.9227 & 1.1452 & 1.1815 & 1.9931 & 0.2588 & 0.3466 & 0.5883 & 0.8403 & 0.8613 & 0.8748 \\
& Keep the Core-Gaussian Noise & 0.9077\textcolor{red}{$\uparrow$} & 0.9164\textcolor{red}{$\uparrow$} & 0.9228\textcolor{red}{$\uparrow$} & 1.1019\textcolor{red}{$\uparrow$} & 1.1495\textcolor{red}{$\uparrow$} & 2.0568\textcolor{green}{$\downarrow$} & 0.2571\textcolor{red}{$\uparrow$} & 0.3462\textcolor{red}{$\uparrow$} & 0.5788\textcolor{red}{$\uparrow$} & 0.8463\textcolor{red}{$\uparrow$} & 0.8643\textcolor{red}{$\uparrow$} & 0.8749\textcolor{red}{$\uparrow$} \\
& Cutout & 0.9081 & 0.9132 & 0.9212 & 1.1342 & 1.2213 & 2.1083 & 0.2667 & 0.3675 & 0.6290 & 0.8472 & 0.8594 & 0.8719 \\
& Keep the Core-Cutout & 0.9099\textcolor{red}{$\uparrow$} & 0.9135\textcolor{red}{$\uparrow$} & 0.9293\textcolor{red}{$\uparrow$} & 1.1006\textcolor{red}{$\uparrow$} & 1.1940\textcolor{red}{$\uparrow$} & 1.9644\textcolor{red}{$\uparrow$} & 0.2624\textcolor{red}{$\uparrow$} & 0.3618\textcolor{red}{$\uparrow$} & 0.5620\textcolor{red}{$\uparrow$} & 0.8498\textcolor{red}{$\uparrow$} & 0.8596\textcolor{red}{$\uparrow$} & 0.8797\textcolor{red}{$\uparrow$} \\

\cmidrule[1pt]{1-14}

\multirow{5}{*}{\textbf{SwinUNet}}
& Baseline & 0.9051 & 0.9145 & 0.9226 & 1.1233 & 1.1778 & 1.7979 & 0.2966 & 0.3501 & 0.5340 & 0.8425 & 0.8619 & 0.8743 \\
& Random Erasing & 0.8989 & 0.9085 & 0.9150 & 1.1831 & 1.2965 & 2.1528 & 0.3351 & 0.3998 & 0.6200 & 0.8316 & 0.8516 & 0.8622 \\
& Keep the Core-Random Erasing & 0.9033\textcolor{red}{$\uparrow$} & 0.9124\textcolor{red}{$\uparrow$} & 0.9193\textcolor{red}{$\uparrow$} & 1.1349\textcolor{red}{$\uparrow$} & 1.2193\textcolor{red}{$\uparrow$} & 1.9367\textcolor{red}{$\uparrow$} & 0.3033\textcolor{red}{$\uparrow$} & 0.3756\textcolor{red}{$\uparrow$} & 0.5915\textcolor{red}{$\uparrow$} & 0.8388\textcolor{red}{$\uparrow$} & 0.8580\textcolor{red}{$\uparrow$} & 0.8692\textcolor{red}{$\uparrow$} \\
& Cutmix & 0.9097 & 0.9182 & 0.9258 & 1.1309 & 1.1685 & 1.6913 & 0.2755 & 0.3299 & 0.4516 & 0.8513 & 0.8688 & 0.8796 \\
& Keep the Core-Cutmix & 0.9113\textcolor{red}{$\uparrow$} & 0.9187\textcolor{red}{$\uparrow$} & 0.9230\textcolor{green}{$\downarrow$} & 1.1297\textcolor{red}{$\uparrow$} & 1.1683\textcolor{red}{$\uparrow$} & 1.6910\textcolor{red}{$\uparrow$} & 0.2748\textcolor{red}{$\uparrow$} & 0.3188\textcolor{red}{$\uparrow$} & 0.4512\textcolor{red}{$\uparrow$} & 0.8517\textcolor{red}{$\uparrow$} & 0.8691\textcolor{red}{$\uparrow$} & 0.8798\textcolor{red}{$\uparrow$} \\

\cmidrule[1pt]{1-14}

\multirow{5}{*}{\textbf{SwinUMamba}}
& Baseline & 0.9207 & 0.9261 & 0.9336 & 1.0219 & 1.0675 & 1.5723 & 0.1999 & 0.2471 & 0.3405 & 0.8688 & 0.8822 & 0.8938 \\
& ABD & 0.9233 & 0.9276 & 0.9362 & 1.0215 & 1.0591 & 1.5057 & 0.1997 & 0.2450 & 0.3426 & 0.8731 & 0.8843 & 0.8972 \\
& Keep the Core-ABD & 0.9234\textcolor{red}{$\uparrow$} & 0.9273\textcolor{green}{$\downarrow$} & 0.9363\textcolor{red}{$\uparrow$} & 1.0218\textcolor{green}{$\downarrow$} & 1.0590\textcolor{red}{$\uparrow$} & 1.4861\textcolor{red}{$\uparrow$} & 0.1996\textcolor{red}{$\uparrow$} & 0.2449\textcolor{red}{$\uparrow$} & 0.3352\textcolor{red}{$\uparrow$} & 0.8739\textcolor{red}{$\uparrow$} & 0.8846\textcolor{red}{$\uparrow$} & 0.8976\textcolor{red}{$\uparrow$} \\
& Mixup & 0.9223 & 0.9310 & 0.9367 & 1.0480 & 1.1090 & 1.6076 & 0.1966 & 0.2346 & 0.3553 & 0.8727 & 0.8908 & 0.9000 \\
& Keep the Core-Mixup & 0.9228\textcolor{red}{$\uparrow$} & 0.9314\textcolor{red}{$\uparrow$} & 0.9369\textcolor{red}{$\uparrow$} & 1.0479\textcolor{red}{$\uparrow$} & 1.1079\textcolor{red}{$\uparrow$} & 1.6070\textcolor{red}{$\uparrow$} & 0.1965\textcolor{red}{$\uparrow$} & 0.2342\textcolor{red}{$\uparrow$} & 0.3530\textcolor{red}{$\uparrow$} & 0.8730\textcolor{red}{$\uparrow$} & 0.8996\textcolor{red}{$\uparrow$} & 0.9040\textcolor{red}{$\uparrow$} \\

\cmidrule[1pt]{1-14}
\end{tabular}}
\label{tab:oasis1_comprehensive_kc_comparison}
\end{table*}

\begin{table*}[t!]
\centering
\caption{Comprehensive evaluation on MRBrainS13 dataset. 'Keep the Core' (KC) methods are compared to their non-KC counterparts (e.g., KC-Gamma vs. Gamma Correction). Metric improvements are marked with a top-right red arrow (\textcolor{red}{$\uparrow$}), degradations with a bottom-right green arrow (\textcolor{green}{$\downarrow$}).}
\resizebox{\textwidth}{!}{%
\renewcommand{\arraystretch}{0.8}
\begin{tabular}{cccccccccccccc}
\cmidrule[1.5pt]{1-14}
\multirow{2}{*}{\textbf{Backbone}} & \multirow{2}{*}{\textbf{Augmentation}} & \multicolumn{3}{c}{\textbf{Dice} $\uparrow$} & \multicolumn{3}{c}{\textbf{HD95(mm)}$\downarrow$} & \multicolumn{3}{c}{\textbf{ASD(mm)}$\downarrow$} & \multicolumn{3}{c}{\textbf{IOU}$\uparrow$} \\
\cmidrule[1pt]{3-14}
& & CSF & GM & WM & CSF & GM & WM & CSF & GM & WM & CSF & GM & WM \\
\cmidrule[1pt]{1-14}

\multirow{5}{*}{\textbf{UNet}}
& Baseline & $0.6947$ & $0.7104$ & $0.7309$ & $1.9670$ & $2.5323$ & $4.7501$ & $0.4383$ & $0.6621$ & $2.0734$ & $0.6115$ & $0.6367$ & $0.6637$ \\
& Gamma Correction & $0.7034$ & $0.7213$ & $0.7409$ & $1.6832$ & $1.5088$ & $3.6378$ & $0.4286$ & $0.4021$ & $1.3495$ & $0.6237$ & $0.6508$ & $0.6734$ \\
& Keep the Core-Gamma & $0.7039^{\textcolor{red}{\uparrow}}$ & $0.7198_{\textcolor{green}{\downarrow}}$ & $0.7414^{\textcolor{red}{\uparrow}}$ & $1.6382^{\textcolor{red}{\uparrow}}$ & $1.6529_{\textcolor{green}{\downarrow}}$ & $3.6230^{\textcolor{red}{\uparrow}}$ & $0.4188^{\textcolor{red}{\uparrow}}$ & $0.4252_{\textcolor{green}{\downarrow}}$ & $1.2716^{\textcolor{red}{\uparrow}}$ & $0.6244^{\textcolor{red}{\uparrow}}$ & $0.6593^{\textcolor{red}{\uparrow}}$ & $0.6737^{\textcolor{red}{\uparrow}}$ \\
& Brightness/Contrast & $0.7028$ & $0.7210$ & $0.7445$ & $1.7027$ & $1.6027$ & $3.7314$ & $0.4118$ & $0.4021$ & $1.5862$ & $0.6228$ & $0.6507$ & $0.6761$ \\
& Keep the Core-Brightness/Contrast & $0.7075^{\textcolor{red}{\uparrow}}$ & $0.7260^{\textcolor{red}{\uparrow}}$ & $0.7414_{\textcolor{green}{\downarrow}}$ & $1.6739^{\textcolor{red}{\uparrow}}$ & $1.5791^{\textcolor{red}{\uparrow}}$ & $3.0973^{\textcolor{red}{\uparrow}}$ & $0.4102^{\textcolor{red}{\uparrow}}$ & $0.3646^{\textcolor{red}{\uparrow}}$ & $1.0075^{\textcolor{red}{\uparrow}}$ & $0.6211_{\textcolor{green}{\downarrow}}$ & $0.6537^{\textcolor{red}{\uparrow}}$ & $0.6786^{\textcolor{red}{\uparrow}}$ \\

\cmidrule[1pt]{1-14}

\multirow{5}{*}{\textbf{SwinUNet}}
& Baseline & $0.6659$ & $0.6936$ & $0.7098$ & $2.3639$ & $3.4329$ & $5.2567$ & $0.5767$ & $0.9310$ & $2.1417$ & $0.5710$ & $0.6151$ & $0.6341$ \\
& Gaussian Blur & $0.6639$ & $0.6928$ & $0.7001$ & $2.3789$ & $3.0653$ & $7.2002$ & $0.6090$ & $0.7770$ & $3.4630$ & $0.5671$ & $0.6100$ & $0.6215$ \\
& Keep the Core-Gaussian Blur & $0.6639$ & $0.6936^{\textcolor{red}{\uparrow}}$ & $0.7043^{\textcolor{red}{\uparrow}}$ & $2.3546^{\textcolor{red}{\uparrow}}$ & $3.0597^{\textcolor{red}{\uparrow}}$ & $7.0595^{\textcolor{red}{\uparrow}}$ & $0.6025^{\textcolor{red}{\uparrow}}$ & $0.7765^{\textcolor{red}{\uparrow}}$ & $2.7829^{\textcolor{red}{\uparrow}}$ & $0.5669_{\textcolor{green}{\downarrow}}$ & $0.6101^{\textcolor{red}{\uparrow}}$ & $0.6237^{\textcolor{red}{\uparrow}}$ \\
& HSMix & $0.6640$ & $0.6917$ & $0.6903$ & $2.5743$ & $3.4378$ & $4.9869$ & $0.6158$ & $1.0094$ & $1.9139$ & $0.5671$ & $0.6087$ & $0.6171$ \\
& Keep the Core-HSMix & $0.6787^{\textcolor{red}{\uparrow}}$ & $0.7063^{\textcolor{red}{\uparrow}}$ & $0.7095^{\textcolor{red}{\uparrow}}$ & $2.1474^{\textcolor{red}{\uparrow}}$ & $1.6678^{\textcolor{red}{\uparrow}}$ & $4.1721^{\textcolor{red}{\uparrow}}$ & $0.5202^{\textcolor{red}{\uparrow}}$ & $0.7576^{\textcolor{red}{\uparrow}}$ & $1.1763^{\textcolor{red}{\uparrow}}$ & $0.5870^{\textcolor{red}{\uparrow}}$ & $0.6282^{\textcolor{red}{\uparrow}}$ & $0.6350^{\textcolor{red}{\uparrow}}$ \\

\cmidrule[1pt]{1-14}

\multirow{5}{*}{\textbf{SwinUMamba}}
& Baseline & $0.6855$ & $0.7007$ & $0.7017$ & $1.9931$ & $3.1604$ & $7.2391$ & $0.5097$ & $0.9757$ & $2.9280$ & $0.5971$ & $0.6211$ & $0.6226$ \\
& MRS & $0.6864$ & $0.7068$ & $0.7103$ & $2.0022$ & $2.4285$ & $5.0634$ & $0.4859$ & $0.6938$ & $1.8649$ & $0.5981$ & $0.6312$ & $0.6368$ \\
& Keep the Core-MRS & $0.7024^{\textcolor{red}{\uparrow}}$ & $0.7207^{\textcolor{red}{\uparrow}}$ & $0.7249^{\textcolor{red}{\uparrow}}$ & $1.7568^{\textcolor{red}{\uparrow}}$ & $1.5814^{\textcolor{red}{\uparrow}}$ & $3.3757^{\textcolor{red}{\uparrow}}$ & $0.4250^{\textcolor{red}{\uparrow}}$ & $0.4365^{\textcolor{red}{\uparrow}}$ & $0.7065^{\textcolor{red}{\uparrow}}$ & $0.6218^{\textcolor{red}{\uparrow}}$ & $0.6499^{\textcolor{red}{\uparrow}}$ & $0.6556^{\textcolor{red}{\uparrow}}$ \\
& Bias Field & $0.6844$ & $0.6995$ & $0.7233$ & $2.0534$ & $3.1019$ & $5.2166$ & $0.5317$ & $0.8232$ & $1.9217$ & $0.5956$ & $0.6218$ & $0.6465$ \\
& Keep the Core-Bias Field & $0.7034^{\textcolor{red}{\uparrow}}$ & $0.7214^{\textcolor{red}{\uparrow}}$ & $0.7525^{\textcolor{red}{\uparrow}}$ & $1.6955^{\textcolor{red}{\uparrow}}$ & $1.5451^{\textcolor{red}{\uparrow}}$ & $3.7876^{\textcolor{red}{\uparrow}}$ & $0.4291^{\textcolor{red}{\uparrow}}$ & $0.3981^{\textcolor{red}{\uparrow}}$ & $1.1525^{\textcolor{red}{\uparrow}}$ & $0.6234^{\textcolor{red}{\uparrow}}$ & $0.6510^{\textcolor{red}{\uparrow}}$ & $0.6780^{\textcolor{red}{\uparrow}}$ \\

\cmidrule[1pt]{1-14}
\end{tabular}}
\label{tab:mrbrains13_comprehensive}
\end{table*}

\subsection{Results on OASIS-1 and MRBrainS13 Datasets}
We conduct a comprehensive evaluation of our "Keep the Core" (KC) framework on two medical image segmentation benchmarks: OASIS-1 and MRBrainS13. As shown in Table~\ref{tab:oasis1_comprehensive_kc_comparison} and Table~\ref{tab:mrbrains13_comprehensive}, we systematically compare standard data augmentation methods against their corresponding "Keep the Core" enhanced versions across three representative backbones: UNet, SwinUNet, and SwinUMamba. From the results, it is observed that while original data augmentation techniques contribute to performance improvements, our proposed "Keep the Core" framework further enhances the efficacy of these SOTA augmentation approaches. Specifically, the KC framework yields substantial performance gains across nearly all augmentation methods.  This is visually evident in both tables, where the red up-arrows (\textcolor{red}{$\uparrow$}) for improvements overwhelmingly outnumber the rare green down-arrows (\textcolor{green}{$\downarrow$}) for metric degradations. On the relatively simpler OASIS-1 dataset (Table~\ref{tab:oasis1_comprehensive_kc_comparison}), the "Keep the Core" framework demonstrates stable improvements. For instance, with the UNet backbone, KC-Gaussian Noise improves the Dice score by 0.39\% (CSF) and 0.23\% (GM) and reduces the HD95 distance by 0.0433mm (CSF) and 0.032mm (GM) compared to the standard Gaussian Noise. On the more advanced SwinUMamba backbone, the KC-ABD method maintains improvements on major metrics with only slight fluctuations in sub-metrics, highlighting the framework's robustness. The performance gains are particularly significant on the more challenging MRBrainS13 dataset (Table~\ref{tab:mrbrains13_comprehensive}). Notably, on the SwinUMamba backbone, the KC-MRS method achieves an absolute Dice score improvement of 1.60\% (CSF) and 1.39\% (GM), while reducing HD95 distance by 0.2454mm (CSF) and 0.8471mm (GM), and ASD by 0.0609mm (CSF) and 1.1584mm (WM). These results indicate our framework's strong adaptability in complex medical imaging scenarios. It is worth noting that our framework's effectiveness is particularly pronounced for information-deleting augmentations (e.g., Cutout, Random Erasing) and mixing-based augmentations (e.g., Mixup, Cutmix). These methods are prone to introducing perturbations, and the significance-preserving mechanism of "Keep the Core" effectively mitigates this issue, leading to sub-gains in overall performance.

\begin{table*}[h!]
\centering
\caption{We evaluate the effectiveness of various data augmentation methods on the MRBrainS13 dataset using importance maps generated by different models. The best-performing metrics in the table are highlighted in bold.}
\resizebox{\textwidth}{!}{%
\renewcommand{\arraystretch}{0.8} 
\begin{tabular}{ccccccccccccc}
\cmidrule[1.5pt](lr){1-13}
\multirow{2}{*}{\textbf{Model}} & \multicolumn{3}{c}{\textbf{Dice} $\uparrow$} & \multicolumn{3}{c}{\textbf{HD95(mm)}$\downarrow$} & \multicolumn{3}{c}{\textbf{ASD(mm)}$\downarrow$} & \multicolumn{3}{c}{\textbf{IOU}$\uparrow$} 
\\ \cmidrule[1pt](lr){2-13}
& CSF & GM & WM 
& CSF & GM & WM 
& CSF & GM & WM 
& CSF & GM & WM 
\\ \cmidrule[1pt](lr){1-13}
SwinUMamba 
& 0.6855 & 0.7007 & 0.7017
& 1.9931 & 3.1604 & 7.2391
& 0.5097 & 0.9757 & 2.9280
& 0.5971 & 0.6211 & 0.6226
\\ \cmidrule[1pt](lr){1-13}
Random Erasing
& 0.6846 & 0.7023 & 0.6997
& 1.9876 & 3.0935 & 6.9380
& 0.5129 & 0.9197 & 3.1028
& 0.5960 & 0.6237 & 0.6239
\\
Keep the core-RandomErasing
& \textbf{0.6999} & \textbf{0.7163} & \textbf{0.7392}
& \textbf{1.7354} & \textbf{1.6958} & \textbf{3.9024}
& \textbf{0.4512} & \textbf{0.4339} & \textbf{1.3344}
& \textbf{0.6180} & \textbf{0.6425} & \textbf{0.6624}
\\
Keep the core(SwinUNet Map)-RandomErasing
& 0.6826 & 0.6960 & 0.7076
& 2.0298 & 2.5886 & 5.8871
& 0.5372 & 0.7075 & 2.1070
& 0.5927 & 0.6136 & 0.6290
\\ \cmidrule[1pt](lr){1-13}
Cutout
& 0.6795 & 0.7000 & 0.6932
& 2.1639 & 3.2561 & 7.2707
& 0.5518 & 1.0129 & 3.0569
& 0.5890 & 0.6204 & 0.6124
\\
Keep the core-Cutout
& \textbf{0.7024} & \textbf{0.7183} & \textbf{0.7369}
& \textbf{1.6590} & \textbf{1.6520} & \textbf{3.7483}
& \textbf{0.4404} & \textbf{0.4769} & \textbf{1.1730}
& \textbf{0.6218} & \textbf{0.6456} & \textbf{0.6597}
\\
Keep the core(SwinUNet Map)-Cutout
& 0.6889 & 0.7097 & 0.7270
& 1.8591 & 3.3772 & 4.2354
& 0.4929 & 0.6573 & 1.6750
& 0.6015 & 0.6327 & 0.6525
\\ \cmidrule[1pt](lr){1-13}
Mixup
& 0.6913 & 0.7090 & 0.7126
& 1.8336 & 2.2244 & \textbf{3.6060}
& 0.5090 & 0.6094 & \textbf{1.0710}
& 0.6049 & 0.6316 & 0.6349
\\
Keep the core-Mixup
& \textbf{0.7033} & \textbf{0.7199} & \textbf{ 0.7353}
& \textbf{1.8056} & \textbf{1.6822} & 3.8079
& \textbf{0.4312} & \textbf{0.3883} & 1.0967
& \textbf{0.6230} & \textbf{0.6486} & \textbf{0.6632}
\\
Keep the core(SwinUNet Map)-Mixup
& 0.6940 & 0.7093 & 0.7275
& 2.0007 & 2.1704 & 5.4621
& 0.5174 & 0.7329 & 2.0954
& 0.5954 & 0.6241 & 0.6414
\\ \cmidrule[1pt](lr){1-13}
Cutmix 
& 0.6808 & 0.6910 & 0.7042
& 2.0764 & 3.4679 & 7.3949
& 0.5312 & 0.9627 & 2.7047
& 0.5925 & 0.6074 & 0.6237
\\
Keep the core-Cutmix
& \textbf{0.6967} & \textbf{0.7130} & 0.7121
& 1.9184 & 2.8017 & \textbf{4.0857}	
& \textbf{0.4494} & 0.7503 & \textbf{1.0156} 
& \textbf{0.6136} & \textbf{0.6398} & 0.6432
\\
Keep the core(SwinUNet Map)-Cutmix
& 0.6875 & 0.7047 & \textbf{0.7311}
& \textbf{1.8900} & \textbf{2.0761} & 4.6900
& 0.5351 & \textbf{0.5630} & 1.7522
& 0.5960 & 0.6254 & \textbf{0.6506}
\\ \cmidrule[1pt](lr){1-13}
\end{tabular}}
\label{tab:ab1}
\end{table*}

\begin{table*}[h!]
\centering
\caption{We evaluate the effectiveness of various data augmentation methods on the MRBrainS13 dataset using importance maps generated by different models. The best-performing metrics in the table are highlighted in bold.}
\resizebox{\textwidth}{!}{%
\renewcommand{\arraystretch}{0.8} 
\begin{tabular}{ccccccccccccc}
\cmidrule[1.5pt](lr){1-13}
\multirow{2}{*}{\textbf{Model}} & \multicolumn{3}{c}{\textbf{Acc} $\uparrow$} & \multicolumn{3}{c}{\textbf{Pre}$\uparrow$} & \multicolumn{3}{c}{\textbf{Sen}$\uparrow$} & \multicolumn{3}{c}{\textbf{Spe}$\uparrow$} 
\\ \cmidrule[1pt](lr){2-13}
& CSF & GM & WM 
& CSF & GM & WM 
& CSF & GM & WM 
& CSF & GM & WM 
\\ \cmidrule[1pt](lr){1-13}
SwinUMamba
& 0.7949 & 0.7912 & 0.9159
& 0.6870 & 0.6854 & 0.7483
& 0.6877 & 0.7215 & 0.6981
& 0.8031 & 0.7986 & 0.9222
\\ \cmidrule[1pt](lr){1-13}
Random Erasing 
& 0.7949 & 0.7916 & \textbf{0.9163}
& \textbf{0.6904} & 0.6917 & 0.7341
& 0.6825 & 0.7187 & 0.7069
& \textbf{0.8035} & 0.7992 & 0.9212
\\
Keep the core-RandomErasing 
& \textbf{0.7966} & \textbf{0.7941} & 0.8967
& 0.6890 & \textbf{0.7107} & \textbf{0.7807}	
& \textbf{0.7129} & \textbf{0.7236} & \textbf{0.7185}
& 0.8027 & \textbf{0.8017} & 0.9023
\\
Keep the core(SwinUNet Map)-RandomErasing 
& 0.7951 & 0.7919 & 0.9157
& 0.6877 & 0.6947 & 0.7489
& 0.7012 & 0.7019 & 0.7022
& 0.8011 & 0.7996 & \textbf{0.9219}
\\ \cmidrule[1pt](lr){1-13}
Cutout
& 0.7940 & 0.7908 & \textbf{0.9158}
& 0.6811 & 0.6818 & 0.7441
& 0.6839 & 0.7254 & 0.6946
& 0.8027 & 0.7977 & \textbf{0.9219}
\\
Keep the core-Cutout
& \textbf{0.7971} & \textbf{0.7944} & 0.8865
& \textbf{0.6970} & \textbf{0.7069} & \textbf{0.7954}
& \textbf{0.7093} & \textbf{0.7317} & 0.7036
& 0.8036 & \textbf{0.8011} & 0.8922
\\
Keep the core(SwinUNet Map)-Cutout
& 0.7956 & 0.7928 & 0.8961
& 0.6933 & 0.6990 & 0.7420
& 0.6809 & 0.7236 & 0\textbf{.7247}
& \textbf{0.8044} & 0.8000 & 0.9006
\\ \cmidrule[1pt](lr){1-13}
Mixup
& 0.7956 & 0.7930 & 0.8752
& 0.6869 & 0.6993 & 0.7488
& 0.6977 & 0.7217 & 0.6922
& 0.8031 & 0.8003 & 0.8809
\\
Keep the core-Mixup 
& \textbf{0.7972} & \textbf{0.7952} & 0.8868
& \textbf{0.6935} & \textbf{0.7193} & \textbf{0.7695}
& \textbf{0.7150} & \textbf{0.7218} & \textbf{0.7190}
& \textbf{0.8032} & \textbf{0.8028} & 0.8915
\\
Keep the core(SwinUNet Map)-Mixup
& 0.7957 & 0.7931 & 0\textbf{.9165}
& 0.6803 & 0.6946 & 0.7616
& 0.6937 & 0.7152 & 0.7151
& 0.8023 & 0.7994 & \textbf{0.9222}
\\ \cmidrule[1pt](lr){1-13}
Cutmix
& 0.7947 & 0.7901 & \textbf{0.9155}
& 0.6819 & 0.6949 & 0.7283
& 0.6863 & 0.6969 & 0.7050
& 0.8027 & 0.7995 & \textbf{0.9211}
\\
Keep the core-Cutmix 
& \textbf{0.7966} & \textbf{0.7936} & 0.8653
& \textbf{0.7009} & \textbf{0.7000} & 0.7496
& \textbf{0.6961} & \textbf{0.7309} & 0.7097
& \textbf{0.8041} & \textbf{0.8005} & 0.8695
\\
Keep the core(SwinUNet Map)-Cutmix
& 0.7948 & 0.7927 & 0.9062
& 0.6800 & 0.6996 & \textbf{0.7606}
& 0.6944 & 0.7123 & \textbf{0.7161}
& 0.8024 & 0.8000 & 0.9123
\\ \cmidrule[1pt](lr){1-13}
\end{tabular}}
\label{tab:ab2}
\end{table*}

\paragraph{Cross-Architecture Consistency}
Our method demonstrates good compatibility across different network architectures. For the relatively simple UNet architecture, the framework provides stable performance gains. The improvements are more pronounced for the Transformer-based SwinUNet. On the state-of-the-art SwinUMamba architecture, the framework still delivers significant benefits. This indicates our method is not limited to a specific architecture and possesses broad applicability. Furthermore, the extent of improvement is influenced by the architectural choice. Compared to simpler CNN architectures, modern architectures with stronger representation capabilities, such as SwinUMamba, can better leverage the saliency priors provided by our framework, thus achieving greater performance enhancements.

\paragraph{Extended Experimental Results \& Qualitative Evaluation}
Tables~\ref{tab:oasis1_comprehensive_kc_comparison} and \ref{tab:mrbrains13_comprehensive} in the main paper present the core comparative results. To provide a more exhaustive validation, we include extensive supplementary tables in the Appendix (e.g., Tables \ref{tab1}-\ref{tab6}). These tables expand upon our main findings, providing a granular breakdown of the systematic comparisons for all 12 augmentation methods across every backbone and metric. This comprehensive dataset serves to rigorously substantiate the effectiveness and robustness of our proposed "Keep the Core" framework. In addition, the corresponding qualitative results are included in the Supplementary Material and are referenced as \cref{OASIS_Results} and \cref{MRBrainS13_Results}.

\subsection{Ablation Study}
\subsubsection{Robustness Evaluation of Importance Maps}
We conduct a systematic comparison of the effects of importance maps on the MRBrainS13 dataset. Specifically, we first use one segmentation model to generate importance maps, and then combine these maps with different data augmentation strategies to train another model, thereby evaluating the robustness of the importance-map generation module in a cross-model setting. As shown in Tables \ref{tab:ab1} and \ref{tab:ab2}, when using the importance maps generated by SwinUNet \cite{swin-unet}, denoted as Keep the Core (SwinUNet Map), to guide data augmentation for SwinUMamba \cite{liu2024swin}, the overall performance is slightly lower than that obtained when using importance maps generated by SwinUMamba itself. Nevertheless, the performance improvement remains stable and significant. These results provide strong additional evidence for the robustness and effectiveness of our proposed offline importance-map generation module, SAGE.

\subsubsection{Feature Discriminability Analysis}
\begin{figure}[t!]
    \centering
    \includegraphics[scale=0.35, trim=0cm 4cm 10cm 8cm, clip]{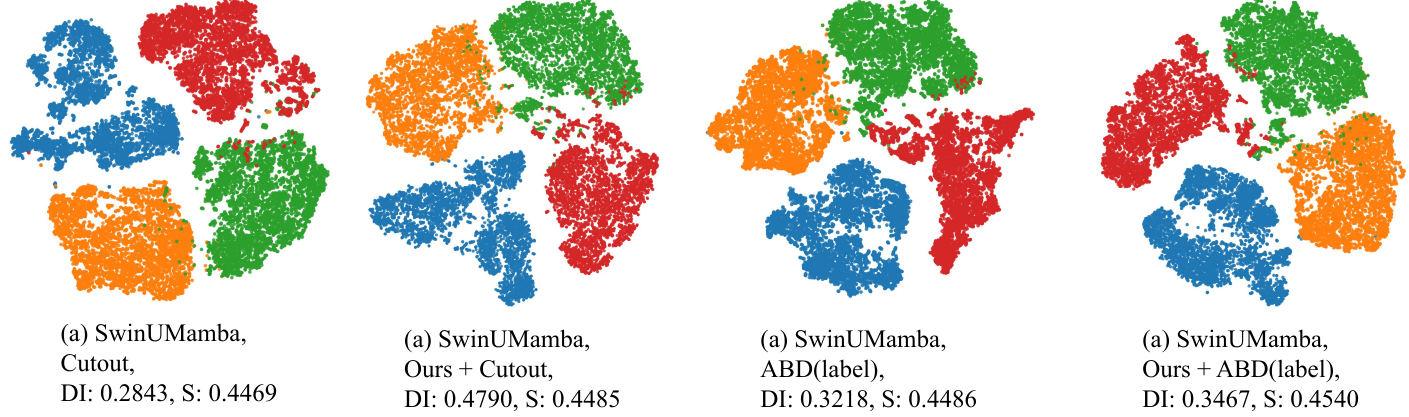}
    \caption{Visualization of deep features on OASIS-1 using t-SNE algorithm \cite{maaten2008visualizing}. Each color denotes a class. DI: Dunn index($\uparrow$). S: Silhouette($\uparrow$, [-1, 1]).}
    \label{fig:t_sne_OASIS}
\end{figure}

\begin{figure}[t!]
    \centering
    \includegraphics[scale=0.35, trim=0cm 12cm 10cm 0cm, clip]{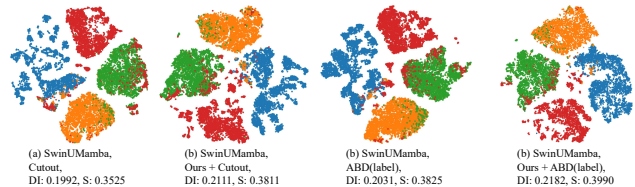}
    \caption{Visualization of deep features on MRBrainS13 using t-SNE algorithm \cite{maaten2008visualizing}. Each color denotes a class. DI: Dunn index($\uparrow$). S: Silhouette($\uparrow$, [-1, 1]).}
    \label{fig:t_sne_MR}
\end{figure}

To intuitively validate our method's efficacy, we visualize the feature space using t-SNE in \cref{fig:t_sne_OASIS} and \cref{fig:t_sne_MR}. The results are definitive. Qualitatively, features from baseline augmentations (e.g., Cutout, ABD) exhibit significant scattering and inter-class overlap. In contrast, our "Keep the Core" variants (Ours + Aug) produce visibly tighter, more compact, and well-defined clusters. This observation is quantitatively confirmed by the substantial improvements in both the Dunn Index (DI $\uparrow$) and Silhouette (S $\uparrow$, [-1, 1]) scores across both datasets. For instance, on OASIS-1, our method boosts the DI for Cutout from 0.2843 to a stark 0.4790. This convergence of visual and metric evidence strongly demonstrates that our framework learns more discriminative representations, successfully minimizing intra-class variance while maximizing inter-class variance. Furthermore, Class Activation Map (CAM) visualizations are provided in the supplementary material (see \cref{fig:OASIS-1_cam} and \cref{fig: MRBrainS13_cam}).

\subsubsection{ The Effect of Parameters}
\begin{figure}[t!]
    \centering
    \begin{subfigure}[t]{0.48\textwidth}
        \centering
        \includegraphics[scale=0.55, trim=0cm 0cm 0cm 0cm, clip]{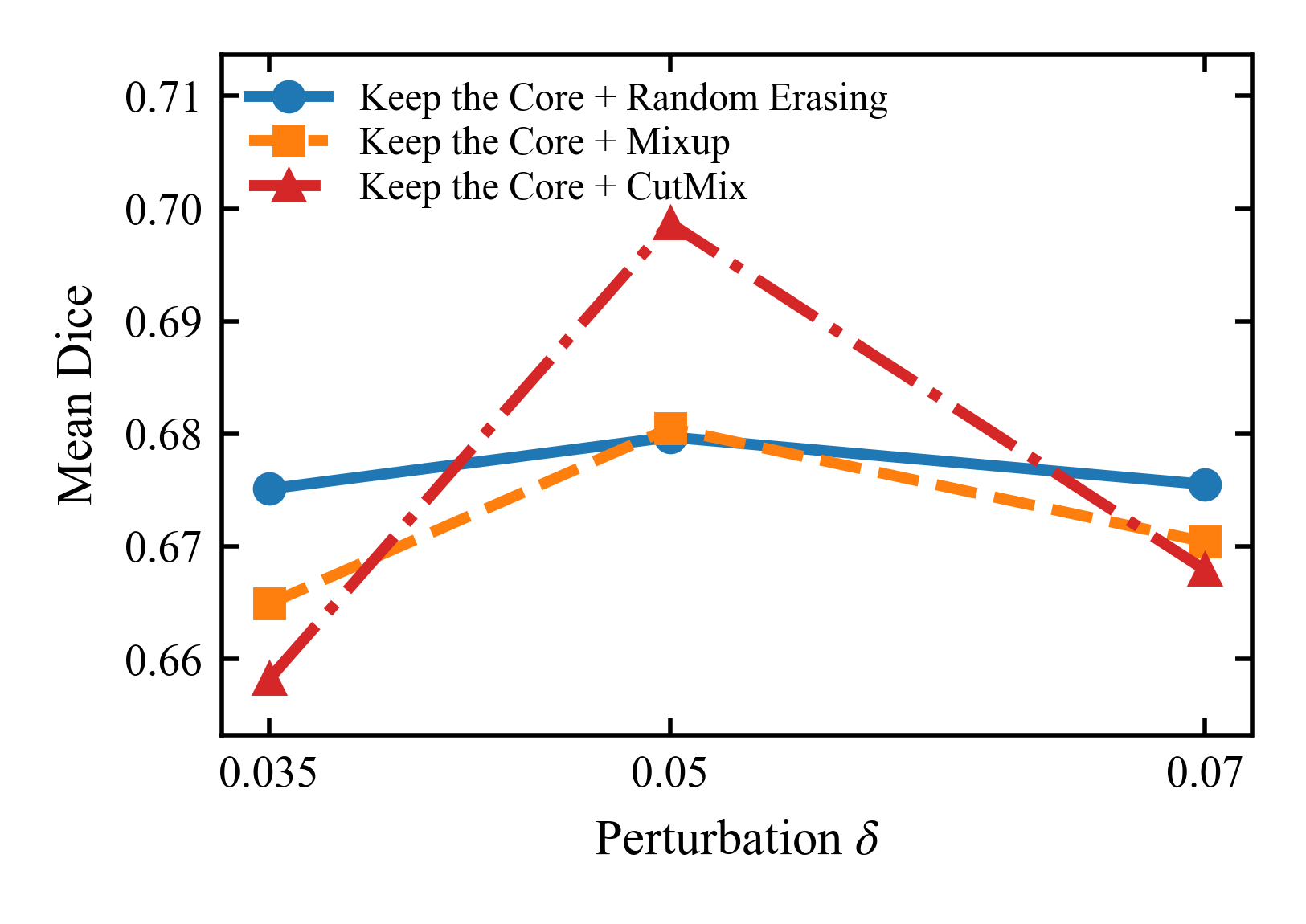}
        \caption{Effect of different perturbation magnitudes on Keep the Core.}
        \label{fig:perturbation}
    \end{subfigure}
    \hfill
    \begin{subfigure}[t]{0.48\textwidth}
        \centering
        \includegraphics[scale=0.55, trim=0cm 0cm 0cm 0cm, clip]{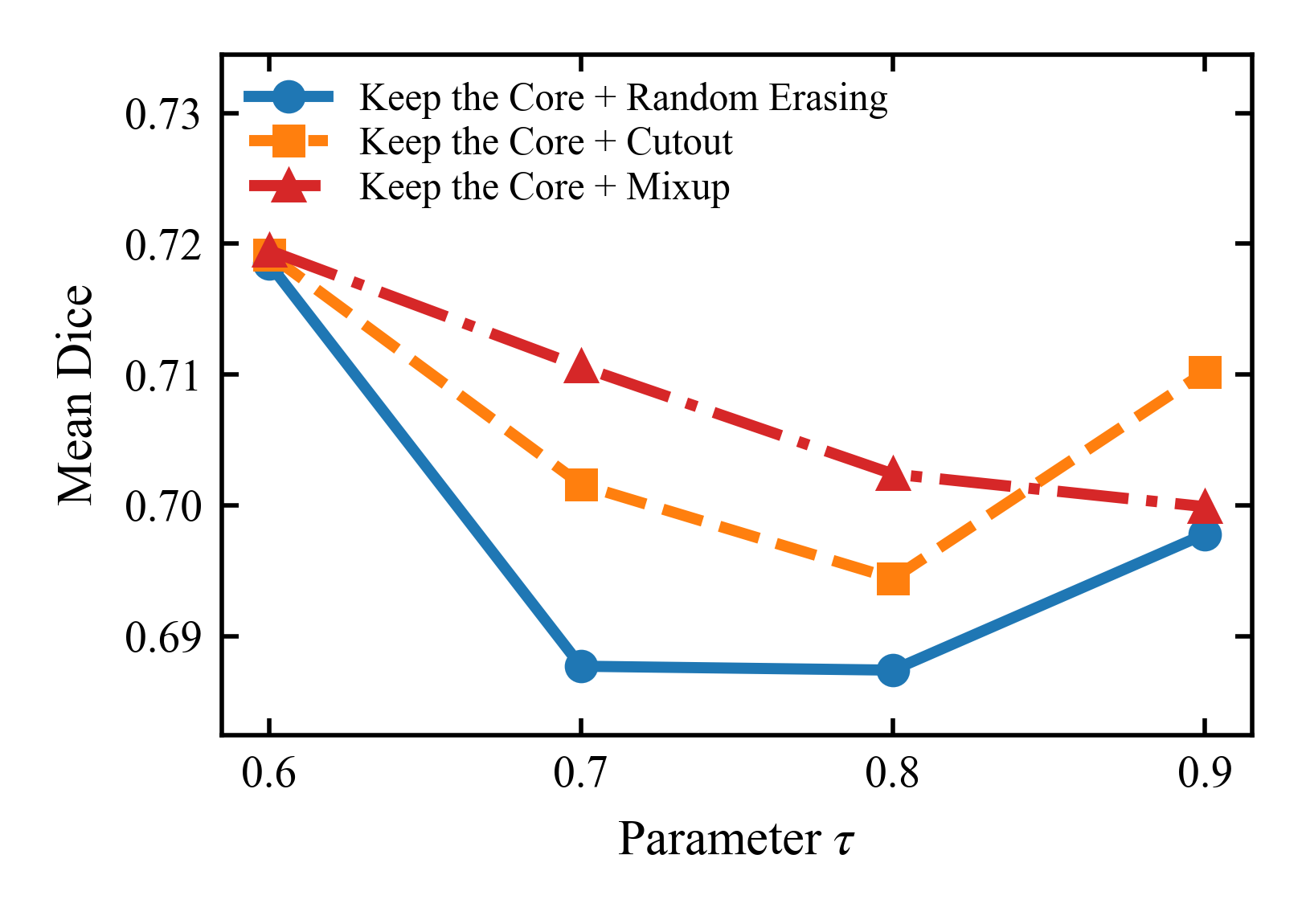}
        \caption{Ablation study on the hyperparameter $\tau$ in Keep the Core.}
        \label{fig:tau}
    \end{subfigure}
    \caption{Ablation Study on Parameter Effectiveness.}
    \label{fig:perturbation_tau}
\end{figure}

To more comprehensively evaluate the effects of different perturbation magnitudes and the hyperparameter $\tau$ on the importance maps, we conduct ablation experiments on the MRBrainS3 dataset, as summarized in \cref{fig:perturbation_tau}. We first combine importance maps generated under different perturbation strengths with multiple data augmentation strategies for testing. As shown in \cref{fig:perturbation_tau} (a), when the perturbation $\delta$ is set to 0.05, the highest average Dice score is achieved consistently across all three augmentation strategies. In addition, \cref{fig:perturbation_tau} (b) shows that Keep the Core attains its best performance when $\tau$ = 60\(\%\). Intuitively, a larger $\tau$ means that the restored regions retain information with higher importance scores. However, an excessively large $\tau$ reduces the effective sample space, thereby limiting the diversity of augmented data and increasing the sampling cost.
 
\section{Conclusion}
\label{sec:conclusion}
In this work, we addressed the feature-agnostic nature of existing augmentation strategies, which often corrupt critical semantics in sparse medical annotations. We proposed "Keep the Core," a novel, data-centric paradigm that redefines feature importance using adversarial vulnerability rather than visual saliency. Our framework introduces SAGE, an offline module that solves a sparse adversarial optimization to identify a minimal set of decision-critical tokens, producing an Importance Map ($W$). This map guides the online KEEP module, which implements a dual-path strategy: (1) "Augment-and-Restore" to strictly preserve high-importance tokens from augmentation-induced corruption, and (2) "Guided Masking" to selectively challenge the model on low-importance tokens. This SAGE-KEEP pipeline effectively unifies and enhances both conventional DA and masked-image modeling. Extensive experiments demonstrate that our backbone-agnostic approach achieves state-of-the-art robustness and generalization on 2D medical segmentation tasks, validating the clear superiority of our significance-preserving paradigm.

\section*{Acknowledgments}
This work was supported by National Natural Science Foundation of China (No.62261053), Tianshan Talent Training Project - Xinjiang Science and Technology Innovation Team Program (2023TSYCTD0012) and Tianshan Innovation Team Program of Xinjiang Uygur Autonomous Region of China (2023D14012).
{
    \small
    \bibliographystyle{ieeenat_fullname}
    \bibliography{main}
}
\clearpage
\setcounter{page}{1}
\maketitlesupplementary

\section{Extended Experimental Results}
In this appendix, we provide the complete and unabridged quantitative results that support the consolidated findings presented in the main paper. The following tables provide an exhaustive, side-by-side comparison of all 12 baseline augmentation methods against their "Keep the Core" (KC) counterparts. The analysis is structured by backbone architecture to demonstrate the universal applicability and scaling advantages of our framework.

\subsection{Deeper Analysis: Validating Our Core Motivation}

Before breaking down results by backbone, we validate our central hypothesis: standard augmentations can be destructive, and "Keep the Core" (KC) mitigates this. We predict two main effects: (1) KC will "rescue" performance from destructive, information-erasing augmentations, and (2) KC will stabilize high-capacity SOTA models on complex datasets.

\paragraph{Evidence for the "Recovery Effect"}
We can see this clearly on the SwinUNet backbone with the OASIS-1 dataset (\cref{tab3}).
\begin{itemize}
    \item \textbf{Gentle Augmentations:} Methods like `Gaussian Noise` (0.9129 $\to$ 0.9131) or `Cutout` (0.9141 $\to$ 0.9144) show minimal gains. The dataset is simple and the augmentations are not overly harmful, so KC has little damage to prevent.
    \item \textbf{Destructive Augmentations:} `Random Erasing` is a clear outlier. Its application causes a significant performance drop (from 0.9145 baseline to 0.9085 GM Dice). This is a perfect example of an "unaware" augmentation corrupting key information. Tellingly, our \textbf{`KC-Random Erasing` (0.9124)} almost entirely recovers this loss, proving that our method is successfully identifying and protecting the salient anatomical features that the baseline method was destroying.
\end{itemize}

\paragraph{Evidence for the "SOTA Stabilization Effect"}
This effect is most evident on the challenging \textbf{MRBrainS13 dataset} with the \textbf{SwinUMamba backbone} (\cref{tab5}). Here, the model is highly complex and the data is difficult.
\begin{itemize}
    \item The baseline model is \textit{extremely sensitive} to augmentation noise. Even "gentle" augmentations like `Gaussian Noise` (0.6961) and `Gaussian Blur` (0.7014) result in GM Dice scores far below the non-augmented baseline (0.7007). The model is clearly confused.
    \item Our KC framework provides a \textit{massive, universal boost}. `KC-Gaussian Noise` jumps to 0.7220, and `KC-Gaussian Blur` jumps to 0.7217.
    \item This pattern holds for \textit{every single augmentation}. The KC variants are not just better; they are in a different performance class entirely. This proves our hypothesis: for SOTA models on complex tasks, a significance-preserving framework is not just a "tweak" but a \textit{necessity} to stabilize training and prevent the model from learning from noise.
\end{itemize}
With this validated motivation, we now analyze the per-backbone results in detail.

\subsection{Analysis for UNet Backbone (Tables \ref{tab1} \& \ref{tab2})}

As shown in \cref{tab1} and \cref{tab2}, the UNet backbone, representing a classic CNN architecture, serves as a fundamental baseline. On the relatively simple \textbf{OASIS-1 dataset}, our KC framework provides consistent and clear improvements. For instance, KC-Cutout improves the baseline Cutout across all primary metrics, notably boosting WM Dice from 0.9212 to 0.9293 and reducing HD95 from 2.1083mm to 1.9644mm.

The benefits are substantially more pronounced on the more challenging \textbf{MRBrainS13 dataset}. Here, standard augmentations sometimes offer marginal gains or even degrade performance, whereas the KC variants provide robust enhancements. A clear example is KC-MRS, which improves Dice (GM) from 0.7172 to 0.7210 and dramatically cuts the ASD (WM) from 1.9747mm down to 1.5167mm. Similarly, KC-ABD achieves a large reduction in HD95 (WM) from 4.3564mm to 3.4333mm. The supplementary metrics in \cref{tab2} (Accuracy, Precision, Sensitivity, Specificity) corroborate these findings, showing holistic performance gains.

\subsection{Analysis for SwinUNet Backbone (Tables \ref{tab3} \& \ref{tab4})}

Upon transitioning to the Transformer-based SwinUNet backbone (\cref{tab3} and \cref{tab4}), we observe that the baseline performance is stronger, yet our KC framework continues to deliver significant advantages.

On \textbf{OASIS-1}, the improvements are stable. As discussed in our motivational analysis, the most significant gain comes from \textbf{KC-Random Erasing}, which recovers the performance drop seen in its baseline counterpart, bringing the Dice (GM) from 0.9085 back up to 0.9124.

Once again, the \textbf{MRBrainS13 dataset} highlights the framework's true strength. The KC enhancements are dramatic, especially for mixing-based and erasing-based methods. KC-HSMix, for example, is a star performer: it elevates Dice (GM) from 0.6917 to 0.7063 and slashes the HD95 (GM) from 3.4378mm to an impressive 1.6678mm. Likewise, KC-Cutmix boosts Dice (CSF) from 0.6641 to 0.6744 and reduces ASD (WM) from 2.3117mm to 1.9485mm. This demonstrates our method's critical ability to stabilize and enhance advanced augmentations on complex architectures.

\subsection{Analysis for SwinUMamba Backbone (Tables \ref{tab5} \& \ref{tab6})}

\cref{tab5} and \cref{tab6} detail the results on the state-of-the-art (SOTA) SwinUMamba backbone. This analysis is crucial, as it tests our framework's ability to improve an already highly optimized model.

On \textbf{OASIS-1}, where performance is approaching saturation (baseline Dice scores are already $\sim$0.92-0.93), our KC methods provide consistent, incremental gains that "polish" the SOTA results. For example, KC-MRS raises the WM Dice from 0.9349 to 0.9373 and KC-Cutout raises the CSF Dice from 0.9229 to 0.9262.

The most compelling evidence is found on the \textbf{MRBrainS13 dataset}. As predicted by our "SOTA Stabilization Effect," the baseline SwinUMamba, while strong, is significantly elevated by our framework across the board. The improvements are not incremental; they are substantial.
\begin{itemize}
    \item \textbf{KC-ABD} boosts the WM Dice from 0.6976 to 0.7381. More strikingly, it reduces the corresponding ASD (Average Surface Distance) from 3.0102mm to a mere \textbf{0.4790mm}—a 6.3x reduction.
    \item \textbf{KC-MRS} achieves the highest CSF Dice (0.7024 vs. 0.6864) and GM Dice (0.7207 vs. 0.7068) among its peers.
    \item \textbf{KC-Bias Field} shows a massive jump in WM Dice (0.7233 to 0.7525) and a large drop in WM HD95 (5.2166mm to 3.7876mm).
\end{itemize}

This focus on the distance metrics (ASD/HD95) is critical. A good Dice score means "most" of the segmentation is correct, but a poor HD95/ASD score means the model has severe "outlier" errors at the boundaries. Our method's dramatic improvements here are the strongest possible evidence for our motivation: by protecting the "core" anatomy, we are specifically preserving the complex, difficult-to-segment boundaries that standard augmentations corrupt. These results conclusively demonstrate that our significance-preserving approach is a vital component for unlocking the full potential of SOTA architectures on complex, clinically-relevant data.

\section{Qualitative Evaluation} 
\cref{OASIS_Results} and \cref{MRBrainS13_Results} provide qualitative comparisons on the OASIS-1 and MRBrainS13 datasets, respectively. It is visually evident that the baseline predictions (c) and standard augmentations (e.g., (d), (g), (i)) frequently suffer from noisy artifacts, blurred boundaries, and mis-segmentation of intricate anatomical structures. In stark contrast, our "Ours+" enhanced methods (e.g., (e), (h), (k)) consistently produce segmentation maps that are visibly cleaner, more spatially coherent, and demonstrate significantly sharper adherence to the Ground Truth (b). This superior fidelity is particularly noticeable in the complex sulcal patterns and holds true across all three backbones (rows), confirming that our framework yields more robust and anatomically plausible results.

\section{Class Activation Map (CAM) Analysis} \label{sec:cam_analysis}
Since class activation maps (CAM) visualization can provide visual explanations for understanding model performance, we incorporate CAM visualization as a tool to analyze the discriminative regions used by our models. We compare the performance of baseline augmentation strategies against their "Ours+" enhanced counterparts. \cref{fig:OASIS-1_cam} and \cref{fig: MRBrainS13_cam} illustrate the CAMs generated from models trained on the OASIS-1 and MRBrainS13 datasets. It is visually evident that the baseline models' (e.g., rows 1, 3, 5) regions of interest are often highly diffuse and noisy. They cover irrelevant textures or artifacts and seem distracted from the core anatomical structures, indicating a potential overfitting problem and poor generalization. In contrast, after utilizing our framework, the "Ours+" models (rows 2, 4, 6) are more inclined to locate and highlight the most relevant parts of the target anatomy (e.g., Gray Matter or White Matter). These methods' regions of interest are significantly more focused, structurally coherent, and precisely localized, while ignoring much of the spurious background information. This indicates a better generalization ability. To summarize, these visualizations underscore how our framework surpasses the efficacy of the original augmentation methodologies, forcing the model to learn true, generalizable anatomical representations, which in turn explains the more robust and accurate segmentation performance.

\section{Computational Cost Analysis.} We analyze the computational overhead in \cref{fig:compute cost}.
A key advantage of our Keep the Core framework is that it introduces no additional latency during the online stage, which is the most critical part of practical usage.
As shown in \cref{fig:compute cost}(b), the online inference time of our method remains identical to the baselines on both MRBrainS13 and OASIS-1.
The radar chart in \cref{fig:compute cost}(a) further summarizes computation-related factors—including model size, FLOPs, training and inference times, and throughput—showing that our modifications do not impose any extra online computation burden.
Although our approach requires an offline preprocessing step to generate importance maps, this cost is paid once per dataset and is fully amortized across experiments, leaving the practical training and inference budgets unchanged.

\begin{table*}[t!]
\centering
\caption{Quantitative comparison of different methods under the same backbone(UNet) on the OASIS-1 and MRBrainS13 test sets. In the table, the best-performing metrics of our method and its corresponding baseline are highlighted in bold.}
\resizebox{\textwidth}{!}{%
\renewcommand{\arraystretch}{0.8} 
\begin{tabular}{ccccccccccccc}
\cmidrule[1.5pt](lr){1-13}
\multirow{2}{*}{\textbf{Model}} & \multicolumn{3}{c}{\textbf{Dice} $\uparrow$} & \multicolumn{3}{c}{\textbf{HD95(mm)}$\downarrow$} & \multicolumn{3}{c}{\textbf{ASD(mm)}$\downarrow$} & \multicolumn{3}{c}{\textbf{IOU}$\uparrow$} 
\\ \cmidrule[1pt](lr){2-13}
& CSF & GM & WM 
& CSF & GM & WM 
& CSF & GM & WM 
& CSF & GM & WM 
\\ \cmidrule[1pt](lr){1-13}
\multicolumn{1}{c}{\textbf{OASIS-1 dataset}} 
\\ 
UNet 
& 0.9062 & 0.9134 & 0.9181 
& 1.1267 & 1.2138 & 2.0824 
& 0.2581 & 0.3611 & 0.6198
& 0.8442 & 0.8599 & 0.8688 
\\ \cmidrule[1pt](lr){1-13}
Gaussian noise
& 0.9038 & 0.9141 & 0.9227
& 1.1452 & 1.1815 & \textbf{1.9931}
& 0.2588 & 0.3466 & 0.5883 
& 0.8403 & 0.8613 & 0.8748
\\ 
Keep the core-Gaussian noise
& \textbf{0.9077} & \textbf{0.9164} & \textbf{0.9228}
& \textbf{1.1019} & \textbf{1.1495} & 2.0568
& \textbf{0.2571} & \textbf{0.3462} & \textbf{0.5788}
& \textbf{0.8463} & \textbf{0.8643} & \textbf{0.8749}
\\ \cmidrule[1pt](lr){1-13}
Gaussian blur
& 0.9056 & 0.9109 & 0.9195
& 1.1286 & 1.2059 & 2.0272
& 0.2630 & 0.3684 & 0.6007
& 0.8429 & 0.8558 & 0.8693
\\ 
Keep the core-Gaussian blur 
& \textbf{0.9087} & \textbf{0.9139} & \textbf{0.9211}
& \textbf{1.1024} & \textbf{1.1795} & \textbf{2.0216}
& \textbf{0.2574} & \textbf{0.3548} & \textbf{0.6004}
& \textbf{0.8478} & \textbf{0.8600} & \textbf{0.8717}
\\
\end{tabular}}
\label{tab1}
\end{table*}

\begin{table*}[t!]
\centering
\caption{Quantitative comparison of different methods under the same backbone(UNet) on the OASIS-1 and MRBrainS13 test sets. In the table, the best-performing metrics of our method and its corresponding baseline are highlighted in bold.}
\resizebox{\textwidth}{!}{%
\renewcommand{\arraystretch}{0.8} 
\begin{tabular}{ccccccccccccc}
\cmidrule[1.5pt](lr){1-13}
\multirow{2}{*}{\textbf{Model}} & \multicolumn{3}{c}{\textbf{Acc} $\uparrow$} & \multicolumn{3}{c}{\textbf{Pre}$\uparrow$} & \multicolumn{3}{c}{\textbf{Sen}$\uparrow$} & \multicolumn{3}{c}{\textbf{Spe}$\uparrow$} 
\\ \cmidrule[1pt](lr){2-13}
& CSF & GM & WM 
& CSF & GM & WM 
& CSF & GM & WM 
& CSF & GM & WM 
\\ \cmidrule[1pt](lr){1-13}
\multicolumn{1}{c}{\textbf{OASIS-1 dataset}} 
\\ 
UNet 
& 0.9732 & 0.9655 & 0.9844 
& 0.9332 & 0.9155 & 0.9174 
& 0.8838 & 0.9140 & 0.9331
& 0.9791 & 0.9711 & 0.9866 
\\ \cmidrule[1pt](lr){1-13}
Gaussian noise
& 0.9729 & 0.9655 & 0.9851
& 0.9336 & 0.9154 & 0.9245
& 0.8791 & 0.9158 & 0.9326
& 0.9791 & 0.9707 & 0.9876
\\
Keep the core-Gaussian noise
& \textbf{0.9734} & \textbf{0.9659} & \textbf{0.9854}
& \textbf{0.9382} & \textbf{0.9184} & \textbf{0.9314}
& \textbf{0.9003} & \textbf{0.9165} & \textbf{0.9345}
& \textbf{0.9792} & \textbf{0.9713} & \textbf{0.9881}
\\ \cmidrule[1pt](lr){1-13}
Gaussian blur
& 0.9730 & 0.9643 & 0.9840
& 0.9369 & 0.9134 & 0.9104
& 0.8795 & 0.9110 & 0.9422
& 0.9795 & 0.9702 & 0.9857
\\
Keep the core-Gaussian blur
& \textbf{0.9737} & \textbf{0.9658} & \textbf{0.9847}
& \textbf{0.9379} & \textbf{0.9174} & \textbf{0.9135}
& \textbf{0.8874} & \textbf{0.9127} & \textbf{0.9424}
& \textbf{0.9796} & \textbf{0.9718} & \textbf{0.9867}
\\
\end{tabular}}
\label{tab2}
\end{table*}

\begin{table*}[t!]
\centering
\caption{Quantitative comparison of different methods under the same backbone(SwinUNet) on the OASIS-1 and MRBrainS13 test sets. In the table, the best-performing metrics of our method and its corresponding baseline are highlighted in bold.}
\resizebox{\textwidth}{!}{%
\renewcommand{\arraystretch}{0.8} 
\begin{tabular}{ccccccccccccc}
\cmidrule[1.5pt](lr){1-13}
\multirow{2}{*}{\textbf{Model}} & \multicolumn{3}{c}{\textbf{Dice} $\uparrow$} & \multicolumn{3}{c}{\textbf{HD95(mm)}$\downarrow$} & \multicolumn{3}{c}{\textbf{ASD(mm)}$\downarrow$} & \multicolumn{3}{c}{\textbf{IOU}$\uparrow$} 
\\ \cmidrule[1pt](lr){2-13}
& CSF & GM & WM 
& CSF & GM & WM 
& CSF & GM & WM 
& CSF & GM & WM 
\\ \cmidrule[1pt](lr){1-13}
\multicolumn{1}{c}{\textbf{OASIS-1 dataset}} 
\\ 
SwinUNet 
& 0.9051 & 0.9145 & 0.9226 
& 1.1233 & 1.1778 & 1.7979	
& 0.2966 & 0.3501 & 0.5340	
& 0.8425 & 0.8619 & 0.8743
\\ \cmidrule[1pt](lr){1-13}
Gaussian noise
& 0.9041 & 0.9129 & 0.9199
& 1.1360 & 1.2119 & 1.8900
& 0.3062 & 0.3605 & 0.5480	
& 0.8403 & 0.8591 & 0.8703
\\ 
Keep the core-Gaussian noise
& \textbf{0.9043} & \textbf{0.9131} & \textbf{0.9221}	
& \textbf{1.1357} & \textbf{1.2108} & \textbf{1.7789}
& \textbf{0.3061} & \textbf{0.3601} & \textbf{0.5478}
& \textbf{0.8406} & \textbf{0.8596} & \textbf{0.8713}
\\ \cmidrule[1pt](lr){1-13}
Gaussian blur 
& 0.9074 & 0.9153 & 0.9216
& 1.1320 & 1.1877 & 1.7252
& 0.3029 & 0.3499 & 0.4994
& 0.8464 & 0.8629 & 0.8732
\\ 
Keep the core-Gaussian blur 
& \textbf{0.9077} & \textbf{0.9154} & \textbf{0.9218}
& \textbf{1.1297} & \textbf{1.1866} & \textbf{1.7139}
& \textbf{0.3026} & \textbf{0.3483} & \textbf{0.4991}
& \textbf{0.8468} & \textbf{0.8630} & \textbf{0.8735}
\\
\end{tabular}}
\label{tab3}
\end{table*}

\begin{table*}[t!]
\centering
\caption{Quantitative comparison of different methods under the same backbone(SwinUNet) on the OASIS-1 and MRBrainS13 test sets. In the table, the best-performing metrics of our method and its corresponding baseline are highlighted in bold.}
\resizebox{\textwidth}{!}{%
\renewcommand{\arraystretch}{0.8} 
\begin{tabular}{ccccccccccccc}
\cmidrule[1.5pt](lr){1-13}
\multirow{2}{*}{\textbf{Model}} & \multicolumn{3}{c}{\textbf{Acc} $\uparrow$} & \multicolumn{3}{c}{\textbf{Pre}$\uparrow$} & \multicolumn{3}{c}{\textbf{Sen}$\uparrow$} & \multicolumn{3}{c}{\textbf{Spe}$\uparrow$} 
\\ \cmidrule[1pt](lr){2-13}
& CSF & GM & WM 
& CSF & GM & WM 
& CSF & GM & WM 
& CSF & GM & WM 
\\ \cmidrule[1pt](lr){1-13}
\multicolumn{1}{c}{\textbf{OASIS-1 dataset}} 
\\ 
SwinUNet 
& 0.9733 & 0.9659 & 0.9853	
& 0.9176 & 0.9168 & 0.9338
& 0.8968 & 0.9159 & 0.9244
& 0.9780 & 0.9716 & 0.9890 
\\ \cmidrule[1pt](lr){1-13}
Gaussian noise
& 0.9731 & 0.9653 & 0.9849
& \textbf{0.9154} & 0.9128 & \textbf{0.9350}
& 0.8969 & 0.9165 & 0.9196
& 0.9778 & 0.9707 & 0.9892
\\
Keep the core-Gaussian noise
& \textbf{0.9733} & \textbf{0.9657} & \textbf{0.9850}
& \textbf{0.9157} & \textbf{0.9130} & \textbf{0.9354}
& 0.8969 & \textbf{0.9166} & 0.9197
& \textbf{0.9810} & \textbf{0.9710} & \textbf{0.9893}
\\ \cmidrule[1pt](lr){1-13}
Gaussian blur 
& 0.9735 & 0.9660 & 0.9852
& 0.9115 & 0.9151 & \textbf{0.9389}
& 0.9071 & 0.9189 & 0.9181
& 0.9776 & 0.9712 & 0.9895
\\
Keep the core-Gaussian blur 
& \textbf{0.9738} & \textbf{0.9671} & \textbf{0.9855}
& \textbf{0.9117} & \textbf{0.9152} & \textbf{0.9389}
& \textbf{0.9075} & \textbf{0.9192} & \textbf{0.9188}
& \textbf{0.9777} & \textbf{0.9717} & \textbf{0.9897}
\\
\end{tabular}}
\label{tab4}
\end{table*}

\begin{table*}[t!]
\centering
\caption{Quantitative comparison of different methods built on the same backbone (SwinUMamba, SOTA) on the OASIS-1 and MRBrainS13 test sets. In the table, the best-performing metrics of our method and its corresponding baseline are highlighted in bold.}
\resizebox{\textwidth}{!}{%
\renewcommand{\arraystretch}{0.8} 
\begin{tabular}{ccccccccccccc}
\cmidrule[1.5pt](lr){1-13}
\multirow{2}{*}{\textbf{Model}} & \multicolumn{3}{c}{\textbf{Dice} $\uparrow$} & \multicolumn{3}{c}{\textbf{HD95(mm)}$\downarrow$} & \multicolumn{3}{c}{\textbf{ASD(mm)}$\downarrow$} & \multicolumn{3}{c}{\textbf{IOU}$\uparrow$} 
\\ \cmidrule[1pt](lr){2-13}
& CSF & GM & WM 
& CSF & GM & WM 
& CSF & GM & WM 
& CSF & GM & WM 
\\ \cmidrule[1pt](lr){1-13}
\multicolumn{1}{c}{\textbf{OASIS-1 dataset}} 
\\ 
SwinUMamba 
& 0.9207 & 0.9261 & 0.9336
& 1.0219 & 1.0675 & 1.5723 
& 0.1999 & 0.2471 & 0.3405
& 0.8688 & 0.8822 & 0.8938
\\ \cmidrule[1pt](lr){1-13}
Gaussian noise 
& 0.9209 & 0.9267 & 0.9350
& 1.0188 & 1.0533 & 1.5169
& 0.2115 & 0.2476 & 0.3446
& 0.8696 & 0.8831 & 0.8957
\\ 
Keep the core-Gaussian noise 
& \textbf{0.9224} & \textbf{0.9269} & \textbf{0.9355}
& \textbf{1.0117} & \textbf{1.0530} & \textbf{1.5088}
& \textbf{0.2046} & \textbf{0.2469} & \textbf{0.3401}
& \textbf{0.8720} & \textbf{0.8834} & \textbf{0.8960}
\\ \cmidrule[1pt](lr){1-13}
Gaussian blur 
& 0.9201 & 0.9251 & 0.9340
& 1.0373 & 1.0745 & 1.5269
& 0.2245 & 0.2496 & 0.3482
& 0.8688 & 0.8821 & 0.8944
\\ 
Keep the core-Gaussian blur
& \textbf{0.9202} & \textbf{0.9257} & \textbf{0.9351}
& \textbf{1.0369} & \textbf{1.0742} & \textbf{1.5209}
& \textbf{0.2167} & \textbf{0.2466} & \textbf{0.3379}
& \textbf{0.8692} & \textbf{0.8822} & \textbf{0.8962}
\\
\end{tabular}}
\label{tab5}
\end{table*}

\begin{table*}[t!]
\centering
\caption{ Quantitative comparison of different methods built on the same backbone (SwinUMamba, SOTA) on the OASIS-1 and MRBrainS13 test sets. In the table, the best-performing metrics of our method and its corresponding baseline are highlighted in bold.}
\resizebox{\textwidth}{!}{%
\renewcommand{\arraystretch}{0.8} 
%
        \end{subfigure}
        \vspace{-5mm} 
        \begin{subfigure}[t]{\textwidth}
            \centering
            \includegraphics[trim=80 18 50 140,clip,width=\linewidth]{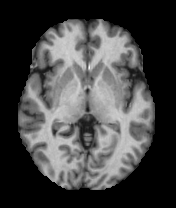}
        \end{subfigure}
        
        \vspace{5mm} 
        \begin{subfigure}[t]{\textwidth}
            \centering
            %
        \end{subfigure}
        \vspace{-5mm} 
        \begin{subfigure}[t]{\textwidth}
            \centering
            \includegraphics[trim=80 18 50 140,clip,width=\linewidth]{sec/fig/OASIS/UNet/2_75_img.png}
        \end{subfigure}
        
        \vspace{5mm} 
        \begin{subfigure}[t]{\textwidth}
            \centering
            %
        \end{subfigure}
        \vspace{-5mm} 
        \begin{subfigure}[t]{\textwidth}
            \centering
            \includegraphics[trim=80 18 50 140,clip,width=\linewidth]{sec/fig/OASIS/UNet/2_75_img.png}
        \end{subfigure} 
    \vspace{1mm}
    \caption{}
    \end{subfigure}
    \hspace{-1.5mm} 
    \begin{subfigure}[t]{0.06\textwidth}
        \centering
        \begin{subfigure}[t]{\textwidth}
            \centering
            %
        \end{subfigure}
        \vspace{-5mm} 
        \begin{subfigure}[t]{\textwidth}
            \centering
            \includegraphics[trim=80 18 50 140,clip,width=\linewidth]{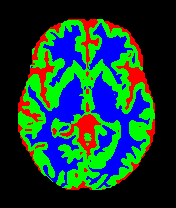}
        \end{subfigure}
        
        \vspace{5mm} 
        \begin{subfigure}[t]{\textwidth}
            \centering
            %
        \end{subfigure}
        \vspace{-5mm} 
        \begin{subfigure}[t]{\textwidth}
            \centering
            \includegraphics[trim=80 18 50 140,,clip,width=\linewidth]{sec/fig/OASIS/UNet/2_75_gt.png}
        \end{subfigure}
        
        \vspace{5mm} 
        \begin{subfigure}[t]{\textwidth}
            \centering
            %
        \end{subfigure}
        \vspace{-5mm} 
        \begin{subfigure}[t]{\textwidth}
            \centering
            \includegraphics[trim=80 18 50 140,clip,width=\linewidth]{sec/fig/OASIS/UNet/2_75_gt.png}
        \end{subfigure}
    \vspace{1mm}
    \subcaption{}
    \end{subfigure}
    \hspace{-1.5mm} 
    \begin{subfigure}[t]{0.06\textwidth}
        \centering
        \begin{subfigure}[t]{\textwidth}
            \centering
            %
        \end{subfigure}
        \vspace{-5mm} 
        \begin{subfigure}[t]{\textwidth}
            \centering
            \includegraphics[trim=80 18 50 140,clip,width=\linewidth]{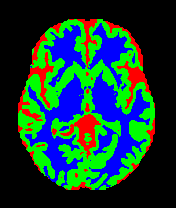}
        \end{subfigure}
        
        \vspace{5mm} 
        \begin{subfigure}[t]{\textwidth}
            \centering
            %
        \end{subfigure}
        \vspace{-5mm} 
        \begin{subfigure}[t]{\textwidth}
            \centering
            \includegraphics[trim=80 18 50 140,,clip,width=\linewidth]{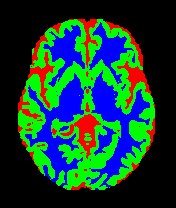}
        \end{subfigure}
        
        \vspace{5mm} 
        \begin{subfigure}[t]{\textwidth}
            \centering
            %
        \end{subfigure}
        \vspace{-5mm} 
        \begin{subfigure}[t]{\textwidth}
            \centering
            \includegraphics[trim=80 18 50 140,clip,width=\linewidth]{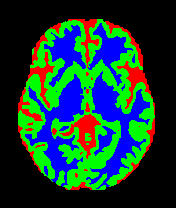}
        \end{subfigure}
    \vspace{1mm}
    \subcaption{}
    \end{subfigure}
    \hspace{-1.5mm} 
    \begin{subfigure}[t]{0.06\textwidth}
        \centering
        \begin{subfigure}[t]{\textwidth}
            \centering
            %
        \end{subfigure}
        \vspace{-5mm} 
        \begin{subfigure}[t]{\textwidth}
            \centering
            \includegraphics[trim=80 18 50 140,clip,width=\linewidth]{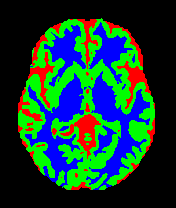}
        \end{subfigure}
        
        \vspace{5mm} 
        \begin{subfigure}[t]{\textwidth}
            \centering
            %
        \end{subfigure}
        \vspace{-5mm} 
        \begin{subfigure}[t]{\textwidth}
            \centering
            \includegraphics[trim=80 18 50 140,,clip,width=\linewidth]{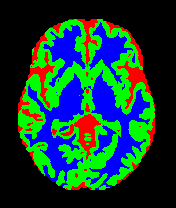}
        \end{subfigure}
        
        \vspace{5mm} 
        \begin{subfigure}[t]{\textwidth}
            \centering
            %
        \end{subfigure}
        \vspace{-5mm} 
        \begin{subfigure}[t]{\textwidth}
            \centering
            \includegraphics[trim=80 18 50 140,clip,width=\linewidth]{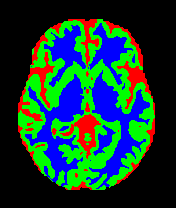}
        \end{subfigure}
    \vspace{1mm}
    \subcaption{}
    \end{subfigure}
    \hspace{-1.5mm} 
    \begin{subfigure}[t]{0.06\textwidth}
        \centering
        \begin{subfigure}[t]{\textwidth}
            \centering
            %
        \end{subfigure}
        \vspace{-5mm} 
        \begin{subfigure}[t]{\textwidth}
            \centering
            \includegraphics[trim=80 18 50 140,clip,width=\linewidth]{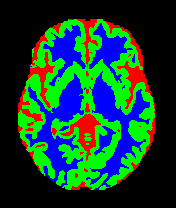}
        \end{subfigure}
        
        \vspace{5mm}
        \begin{subfigure}[t]{\textwidth}
            \centering
            %
        \end{subfigure}
        \vspace{-5mm} 
        \begin{subfigure}[t]{\textwidth}
            \centering
            \includegraphics[trim=80 18 50 140,,clip,width=\linewidth]{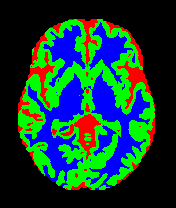}
        \end{subfigure}
        
        \vspace{5mm}
        \begin{subfigure}[t]{\textwidth}
            \centering
            %
        \end{subfigure}
        \vspace{-5mm} 
        \begin{subfigure}[t]{\textwidth}
            \centering
            \includegraphics[trim=80 18 50 140,clip,width=\linewidth]{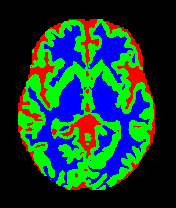}
        \end{subfigure}
    \vspace{1mm}
    \subcaption{}
    \end{subfigure}
    \hspace{-1.5mm} 
    \begin{subfigure}[t]{0.06\textwidth}
        \centering
        \begin{subfigure}[t]{\textwidth}
            \centering
            %
        \end{subfigure}
        \vspace{-5mm} 
        \begin{subfigure}[t]{\textwidth}
            \centering
            \includegraphics[trim=80 18 50 140,clip,width=\linewidth]{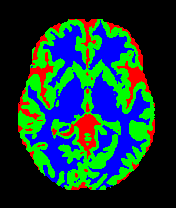}
        \end{subfigure}
        
        \vspace{5mm}
        \begin{subfigure}[t]{\textwidth}
            \centering
            %
        \end{subfigure}
        \vspace{-5mm} 
        \begin{subfigure}[t]{\textwidth}
            \centering
            \includegraphics[trim=80 18 50 140,,clip,width=\linewidth]{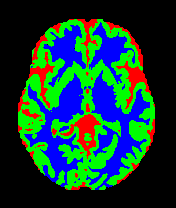}
        \end{subfigure}
        
        \vspace{5mm}
        \begin{subfigure}[t]{\textwidth}
            \centering
            %
        \end{subfigure}
        \vspace{-5mm} 
        \begin{subfigure}[t]{\textwidth}
            \centering
            \includegraphics[trim=80 18 50 140,clip,width=\linewidth]{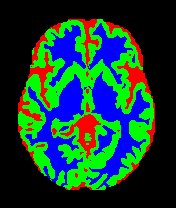}
        \end{subfigure}
    \vspace{1mm}
    \subcaption{}
    \end{subfigure}
    \hspace{-1.5mm} 
    \begin{subfigure}[t]{0.06\textwidth}
        \centering
        \begin{subfigure}[t]{\textwidth}
            \centering
            %
        \end{subfigure}
        \vspace{-5mm} 
        \begin{subfigure}[t]{\textwidth}
            \centering
            \includegraphics[trim=80 18 50 140,clip,width=\linewidth]{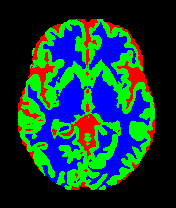}
        \end{subfigure}
        
        \vspace{5mm}
        \begin{subfigure}[t]{\textwidth}
            \centering
            %
        \end{subfigure}
        \vspace{-5mm} 
        \begin{subfigure}[t]{\textwidth}
            \centering
            \includegraphics[trim=80 18 50 140,,clip,width=\linewidth]{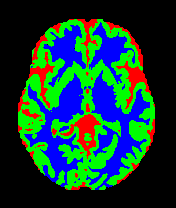}
        \end{subfigure}
        
        \vspace{5mm}
        \begin{subfigure}[t]{\textwidth}
            \centering
            %
        \end{subfigure}
        \vspace{-5mm} 
        \begin{subfigure}[t]{\textwidth}
            \centering
            \includegraphics[trim=80 18 50 140,clip,width=\linewidth]{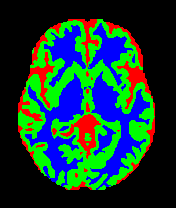}
        \end{subfigure}
    \vspace{1mm}
    \subcaption{}
    \end{subfigure}
    \hspace{-1.5mm} 
    \begin{subfigure}[t]{0.06\textwidth}
        \centering
        \begin{subfigure}[t]{\textwidth}
            \centering
            %
        \end{subfigure}
        \vspace{-5mm} 
        \begin{subfigure}[t]{\textwidth}
            \centering
            \includegraphics[trim=80 18 50 140,clip,width=\linewidth]{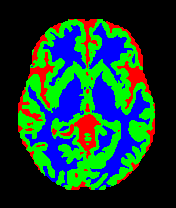}
        \end{subfigure}
        
        \vspace{5mm}
        \begin{subfigure}[t]{\textwidth}
            \centering
            %
        \end{subfigure}
        \vspace{-5mm} 
        \begin{subfigure}[t]{\textwidth}
            \centering
            \includegraphics[trim=80 18 50 140,,,clip,width=\linewidth]{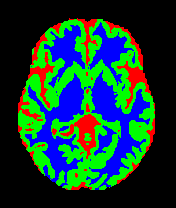}
        \end{subfigure}
        
        \vspace{5mm}
        \begin{subfigure}[t]{\textwidth}
            \centering
            %
        \end{subfigure}
        \vspace{-5mm} 
        \begin{subfigure}[t]{\textwidth}
            \centering
            \includegraphics[trim=80 18 50 140,clip,width=\linewidth]{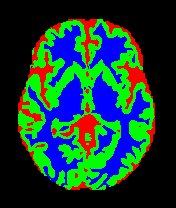}
        \end{subfigure}
    \vspace{1mm}
    \subcaption{}
    \end{subfigure}
    \hspace{-1.5mm}
    \begin{subfigure}[t]{0.06\textwidth}
        \centering
        \begin{subfigure}[t]{\textwidth}
            \centering
            %
        \end{subfigure}
        \vspace{-5mm} 
        \begin{subfigure}[t]{\textwidth}
            \centering
            \includegraphics[trim=80 18 50 140,clip,width=\linewidth]{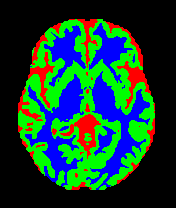}
        \end{subfigure}
        
        \vspace{5mm}
        \begin{subfigure}[t]{\textwidth}
            \centering
            %
        \end{subfigure}
        \vspace{-5mm} 
        \begin{subfigure}[t]{\textwidth}
            \centering
            \includegraphics[trim=80 18 50 140,,clip,width=\linewidth]{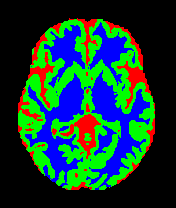}
        \end{subfigure}
        
        \vspace{5mm}
        \begin{subfigure}[t]{\textwidth}
            \centering
            %
        \end{subfigure}
        \vspace{-5mm} 
        \begin{subfigure}[t]{\textwidth}
            \centering
            \includegraphics[trim=80 18 50 140,clip,width=\linewidth]{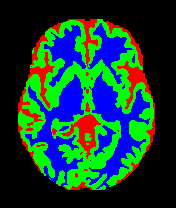}
        \end{subfigure}
    \vspace{1mm}
    \subcaption{}
    \end{subfigure}
    \hspace{-1.5mm}
    \begin{subfigure}[t]{0.06\textwidth}
        \centering
        \begin{subfigure}[t]{\textwidth}
            \centering
            %
        \end{subfigure}
        \vspace{-5mm} 
        \begin{subfigure}[t]{\textwidth}
            \centering
            \includegraphics[trim=80 18 50 140,clip,width=\linewidth]{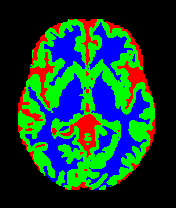}
        \end{subfigure}
        
        \vspace{5mm}
        \begin{subfigure}[t]{\textwidth}
            \centering
            %
        \end{subfigure}
        \vspace{-5mm} 
        \begin{subfigure}[t]{\textwidth}
            \centering
            \includegraphics[trim=80 18 50 140,,clip,width=\linewidth]{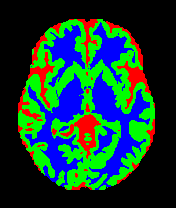}
        \end{subfigure}
        
        \vspace{5mm}
        \begin{subfigure}[t]{\textwidth}
            \centering
            %
        \end{subfigure}
        \vspace{-5mm} 
        \begin{subfigure}[t]{\textwidth}
            \centering
            \includegraphics[trim=80 18 50 140,clip,width=\linewidth]{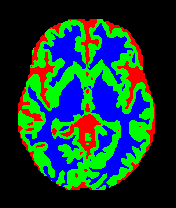}
        \end{subfigure}
    \vspace{1mm}
    \subcaption{}
    \end{subfigure}
    \hspace{-1.5mm}
    \begin{subfigure}[t]{0.06\textwidth}
        \centering
        \begin{subfigure}[t]{\textwidth}
            \centering
            %
        \end{subfigure}
        \vspace{-5mm} 
        \begin{subfigure}[t]{\textwidth}
            \centering
            \includegraphics[trim=80 18 50 140,clip,width=\linewidth]{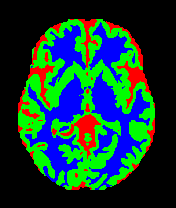}
        \end{subfigure}
        
        \vspace{5mm}
        \begin{subfigure}[t]{\textwidth}
            \centering
            %
        \end{subfigure}
        \vspace{-5mm} 
        \begin{subfigure}[t]{\textwidth}
            \centering
            \includegraphics[trim=80 18 50 140,,clip,width=\linewidth]{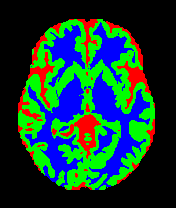}
        \end{subfigure}
        
        \vspace{5mm}
        \begin{subfigure}[t]{\textwidth}
            \centering
            %
        \end{subfigure}
        \vspace{-5mm} 
        \begin{subfigure}[t]{\textwidth}
            \centering
            \includegraphics[trim=80 18 50 140,clip,width=\linewidth]{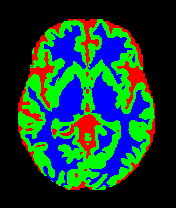}
        \end{subfigure}
    \vspace{1mm}
    \subcaption{}
    \end{subfigure}
    \hspace{-1.5mm}
\caption{Comparison of segmentation results on the OASIS-1 dataset under different combinations of models and augmentation strategies. From top to bottom, the rows correspond to UNet, SwinUNet, and SwinUMamba. (a) Image, (b) GT, (c) Pred, (d) Gaussian noise, (e) Ours+Gaussian noise, (f) Mixup, (g) Ours+Mixup, (h) Cutmix, (i) Ours+Cutmix, (j) MRS(label), (k) Ours+MRS(label).}
\label{OASIS_Results}
\end{figure*}

\begin{figure*}[t!]
    \centering
    \begin{subfigure}[t]{0.06\textwidth}
        \centering
        \begin{subfigure}[t]{\textwidth}
            \centering

        \end{subfigure}
        \vspace{-5mm} 
        \begin{subfigure}[t]{\textwidth}
            \centering
            \includegraphics[trim=80 18 50 140,clip,width=\linewidth]{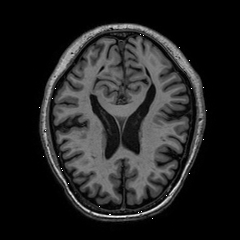}
        \end{subfigure}
        
        \vspace{5mm} 
        \begin{subfigure}[t]{\textwidth}
            \centering
            %
        \end{subfigure}
        \vspace{-5mm} 
        \begin{subfigure}[t]{\textwidth}
            \centering
            \includegraphics[trim=80 18 50 140,clip,width=\linewidth]{sec/fig/MRBrainS13/SwinUMamba/15_26_img.png}
        \end{subfigure}
        
        \vspace{5mm} 
        \begin{subfigure}[t]{\textwidth}
            \centering
            %
        \end{subfigure}
        \vspace{-5mm} 
        \begin{subfigure}[t]{\textwidth}
            \centering
            \includegraphics[trim=80 18 50 140,clip,width=\linewidth]{sec/fig/MRBrainS13/SwinUMamba/15_26_img.png}
        \end{subfigure} 
    \vspace{1mm}
    \caption{}
    \end{subfigure}
    \hspace{-1.5mm} 
    \begin{subfigure}[t]{0.06\textwidth}
        \centering
        \begin{subfigure}[t]{\textwidth}
            \centering
            %
        \end{subfigure}
        \vspace{-5mm} 
        \begin{subfigure}[t]{\textwidth}
            \centering
            \includegraphics[trim=80 18 50 140,clip,width=\linewidth]{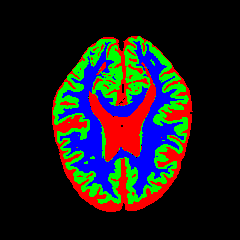}
        \end{subfigure}
        
        \vspace{5mm} 
        \begin{subfigure}[t]{\textwidth}
            \centering
            %
        \end{subfigure}
        \vspace{-5mm} 
        \begin{subfigure}[t]{\textwidth}
            \centering
            \includegraphics[trim=80 18 50 140,clip,width=\linewidth]{sec/fig/MRBrainS13/SwinUMamba/15_26_gt.png}
        \end{subfigure}
        
        \vspace{5mm} 
        \begin{subfigure}[t]{\textwidth}
            \centering
            %
        \end{subfigure}
        \vspace{-5mm} 
        \begin{subfigure}[t]{\textwidth}
            \centering
            \includegraphics[trim=80 18 50 140,clip,width=\linewidth]{sec/fig/MRBrainS13/SwinUMamba/15_26_gt.png}
        \end{subfigure}
    \vspace{1mm}
    \subcaption{}
    \end{subfigure}
    \hspace{-1.5mm} 
    \begin{subfigure}[t]{0.06\textwidth}
        \centering
        \begin{subfigure}[t]{\textwidth}
            \centering
            %
        \end{subfigure}
        \vspace{-5mm} 
        \begin{subfigure}[t]{\textwidth}
            \centering
            \includegraphics[trim=80 18 50 140,clip,width=\linewidth]{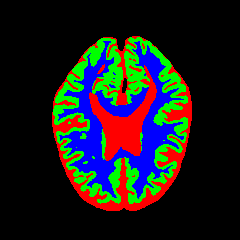}
        \end{subfigure}
        
        \vspace{5mm} 
        \begin{subfigure}[t]{\textwidth}
            \centering
            %
        \end{subfigure}
        \vspace{-5mm} 
        \begin{subfigure}[t]{\textwidth}
            \centering
            \includegraphics[trim=80 18 50 140,clip,width=\linewidth]{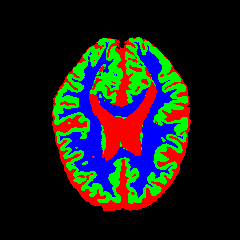}
        \end{subfigure}
        
        \vspace{5mm} 
        \begin{subfigure}[t]{\textwidth}
            \centering
            %
        \end{subfigure}
        \vspace{-5mm} 
        \begin{subfigure}[t]{\textwidth}
            \centering
            \includegraphics[trim=80 18 50 140,clip,width=\linewidth]{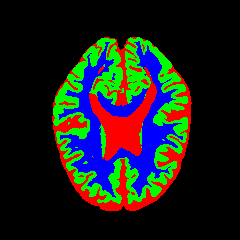}
        \end{subfigure}
    \vspace{1mm}
    \subcaption{}
    \end{subfigure}
    \hspace{-1.5mm} 
    \begin{subfigure}[t]{0.06\textwidth}
        \centering
        \begin{subfigure}[t]{\textwidth}
            \centering
            %
        \end{subfigure}
        \vspace{-5mm} 
        \begin{subfigure}[t]{\textwidth}
            \centering
            \includegraphics[trim=80 18 50 140,clip,width=\linewidth]{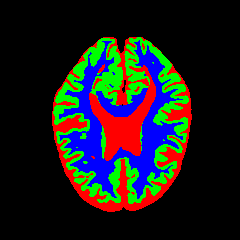}
        \end{subfigure}
        
        \vspace{5mm} 
        \begin{subfigure}[t]{\textwidth}
            \centering
            %
        \end{subfigure}
        \vspace{-5mm} 
        \begin{subfigure}[t]{\textwidth}
            \centering
            \includegraphics[trim=80 18 50 140,clip,width=\linewidth]{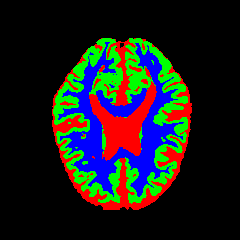}
        \end{subfigure}
        
        \vspace{5mm} 
        \begin{subfigure}[t]{\textwidth}
            \centering
            %
        \end{subfigure}
        \vspace{-5mm} 
        \begin{subfigure}[t]{\textwidth}
            \centering
            \includegraphics[trim=80 18 50 140,clip,width=\linewidth]{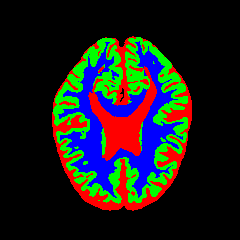}
        \end{subfigure}
    \vspace{1mm}
    \subcaption{}
    \end{subfigure}
    \hspace{-1.5mm} 
    \begin{subfigure}[t]{0.06\textwidth}
        \centering
        \begin{subfigure}[t]{\textwidth}
            \centering
            %
        \end{subfigure}
        \vspace{-5mm} 
        \begin{subfigure}[t]{\textwidth}
            \centering
            \includegraphics[trim=80 18 50 140,clip,width=\linewidth]{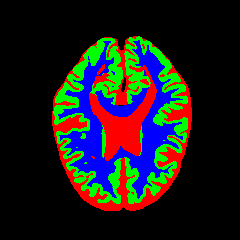}
        \end{subfigure}
        
        \vspace{5mm}
        \begin{subfigure}[t]{\textwidth}
            \centering
            %
        \end{subfigure}
        \vspace{-5mm} 
        \begin{subfigure}[t]{\textwidth}
            \centering
            \includegraphics[trim=80 18 50 140,clip,width=\linewidth]{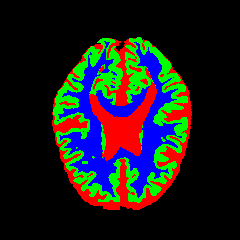}
        \end{subfigure}
        
        \vspace{5mm}
        \begin{subfigure}[t]{\textwidth}
            \centering
            %
        \end{subfigure}
        \vspace{-5mm} 
        \begin{subfigure}[t]{\textwidth}
            \centering
            \includegraphics[trim=80 18 50 140,clip,width=\linewidth]{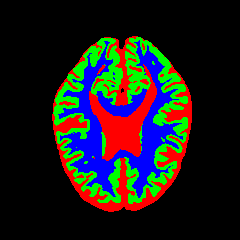}
        \end{subfigure}
    \vspace{1mm}
    \subcaption{}
    \end{subfigure}
    \hspace{-1.5mm} 
    \begin{subfigure}[t]{0.06\textwidth}
        \centering
        \begin{subfigure}[t]{\textwidth}
            \centering
            %
        \end{subfigure}
        \vspace{-5mm} 
        \begin{subfigure}[t]{\textwidth}
            \centering
            \includegraphics[trim=80 18 50 140,clip,width=\linewidth]{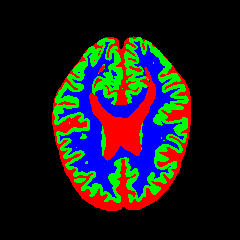}
        \end{subfigure}
        
        \vspace{5mm}
        \begin{subfigure}[t]{\textwidth}
            \centering
            %
        \end{subfigure}
        \vspace{-5mm} 
        \begin{subfigure}[t]{\textwidth}
            \centering
            \includegraphics[trim=80 18 50 140,clip,width=\linewidth]{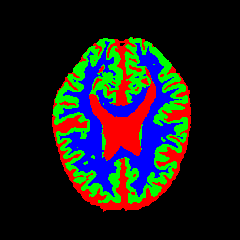}
        \end{subfigure}
        
        \vspace{5mm}
        \begin{subfigure}[t]{\textwidth}
            \centering
            %
        \end{subfigure}
        \vspace{-5mm} 
        \begin{subfigure}[t]{\textwidth}
            \centering
            \includegraphics[trim=80 18 50 140,clip,width=\linewidth]{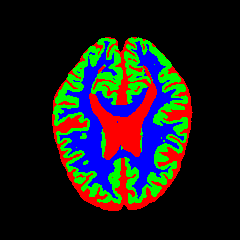}
        \end{subfigure}
    \vspace{1mm}
    \subcaption{}
    \end{subfigure}
    \hspace{-1.5mm} 
    \begin{subfigure}[t]{0.06\textwidth}
        \centering
        \begin{subfigure}[t]{\textwidth}
            \centering
            %
        \end{subfigure}
        \vspace{-5mm} 
        \begin{subfigure}[t]{\textwidth}
            \centering
            \includegraphics[trim=80 18 50 140,clip,width=\linewidth]{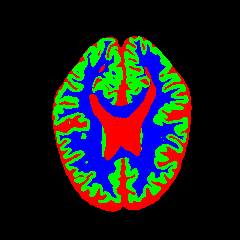}
        \end{subfigure}
        
        \vspace{5mm}
        \begin{subfigure}[t]{\textwidth}
            \centering
            %
        \end{subfigure}
        \vspace{-5mm} 
        \begin{subfigure}[t]{\textwidth}
            \centering
            \includegraphics[trim=80 18 50 140,clip,width=\linewidth]{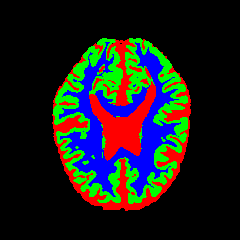}
        \end{subfigure}
        
        \vspace{5mm}
        \begin{subfigure}[t]{\textwidth}
            \centering
            %
        \end{subfigure}
        \vspace{-5mm} 
        \begin{subfigure}[t]{\textwidth}
            \centering
            \includegraphics[trim=80 18 50 140,clip,width=\linewidth]{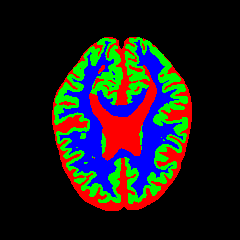}
        \end{subfigure}
    \vspace{1mm}
    \subcaption{}
    \end{subfigure}
    \hspace{-1.5mm} 
    \begin{subfigure}[t]{0.06\textwidth}
        \centering
        \begin{subfigure}[t]{\textwidth}
            \centering
            %
        \end{subfigure}
        \vspace{-5mm} 
        \begin{subfigure}[t]{\textwidth}
            \centering
            \includegraphics[trim=80 18 50 140,clip,width=\linewidth]{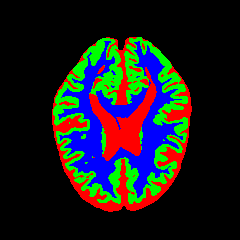}
        \end{subfigure}
        
        \vspace{5mm}
        \begin{subfigure}[t]{\textwidth}
            \centering
            %
        \end{subfigure}
        \vspace{-5mm} 
        \begin{subfigure}[t]{\textwidth}
            \centering
            \includegraphics[trim=80 18 50 140,clip,width=\linewidth]{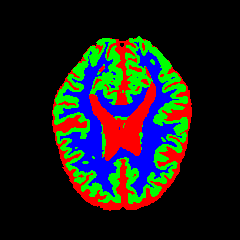}
        \end{subfigure}
        
        \vspace{5mm}
        \begin{subfigure}[t]{\textwidth}
            \centering
            %
        \end{subfigure}
        \vspace{-5mm} 
        \begin{subfigure}[t]{\textwidth}
            \centering
            \includegraphics[trim=80 18 50 140,clip,width=\linewidth]{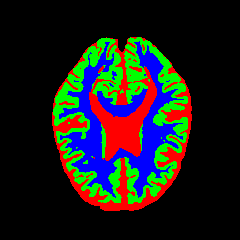}
        \end{subfigure}
    \vspace{1mm}
    \subcaption{}
    \end{subfigure}
    \hspace{-1.5mm}
    \begin{subfigure}[t]{0.06\textwidth}
        \centering
        \begin{subfigure}[t]{\textwidth}
            \centering
            %
        \end{subfigure}
        \vspace{-5mm} 
        \begin{subfigure}[t]{\textwidth}
            \centering
            \includegraphics[trim=80 18 50 140,clip,width=\linewidth]{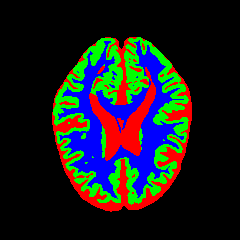}
        \end{subfigure}
        
        \vspace{5mm}
        \begin{subfigure}[t]{\textwidth}
            \centering
            %
        \end{subfigure}
        \vspace{-5mm} 
        \begin{subfigure}[t]{\textwidth}
            \centering
            \includegraphics[trim=80 18 50 140,clip,width=\linewidth]{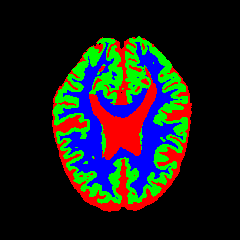}
        \end{subfigure}
        
        \vspace{5mm}
        \begin{subfigure}[t]{\textwidth}
            \centering
            %
        \end{subfigure}
        \vspace{-5mm} 
        \begin{subfigure}[t]{\textwidth}
            \centering
            \includegraphics[trim=80 18 50 140,clip,width=\linewidth]{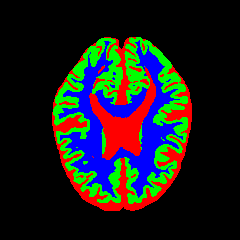}
        \end{subfigure}
    \vspace{1mm}
    \subcaption{}
    \end{subfigure}
    \hspace{-1.5mm}
    \begin{subfigure}[t]{0.06\textwidth}
        \centering
        \begin{subfigure}[t]{\textwidth}
            \centering
            %
        \end{subfigure}
        \vspace{-5mm} 
        \begin{subfigure}[t]{\textwidth}
            \centering
            \includegraphics[trim=80 18 50 140,clip,width=\linewidth]{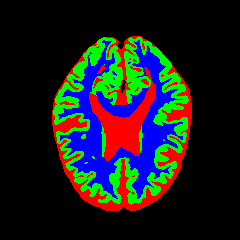}
        \end{subfigure}
        
        \vspace{5mm}
        \begin{subfigure}[t]{\textwidth}
            \centering
            %
        \end{subfigure}
        \vspace{-5mm} 
        \begin{subfigure}[t]{\textwidth}
            \centering
            \includegraphics[trim=80 18 50 140,clip,width=\linewidth]{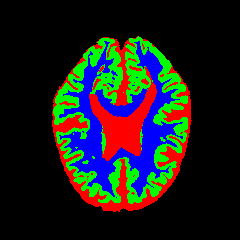}
        \end{subfigure}
        
        \vspace{5mm}
        \begin{subfigure}[t]{\textwidth}
            \centering
            %
        \end{subfigure}
        \vspace{-5mm} 
        \begin{subfigure}[t]{\textwidth}
            \centering
            \includegraphics[trim=80 18 50 140,clip,width=\linewidth]{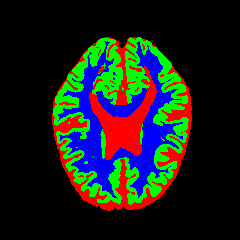}
        \end{subfigure}
    \vspace{1mm}
    \subcaption{}
    \end{subfigure}
    \hspace{-1.5mm}
    \begin{subfigure}[t]{0.06\textwidth}
        \centering
        \begin{subfigure}[t]{\textwidth}
            \centering
            %
        \end{subfigure}
        \vspace{-5mm} 
        \begin{subfigure}[t]{\textwidth}
            \centering
            \includegraphics[trim=80 18 50 140,clip,width=\linewidth]{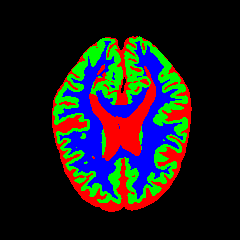}
        \end{subfigure}
        
        \vspace{5mm}
        \begin{subfigure}[t]{\textwidth}
            \centering
            %
        \end{subfigure}
        \vspace{-5mm} 
        \begin{subfigure}[t]{\textwidth}
            \centering
            \includegraphics[trim=80 18 50 140,clip,width=\linewidth]{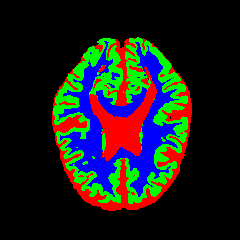}
        \end{subfigure}
        
        \vspace{5mm}
        \begin{subfigure}[t]{\textwidth}
            \centering
            %
        \end{subfigure}
        \vspace{-5mm} 
        \begin{subfigure}[t]{\textwidth}
            \centering
            \includegraphics[trim=80 18 50 140,clip,width=\linewidth]{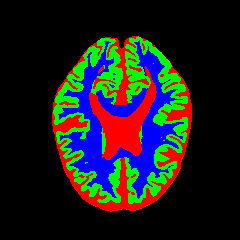}
        \end{subfigure}
    \vspace{1mm}
    \subcaption{}
    \end{subfigure}
    \hspace{-1.5mm}
\caption{Comparison of segmentation results on the MRBrainS13 dataset under different combinations of models and augmentation strategies. From top to bottom, the rows correspond to UNet, SwinUNet, and SwinUMamba. (a) Image, (b) GT, (c) Pred, (d) Gaussian noise, (e) Ours+Gaussian noise, (f) Mixup, (g) Ours+Mixup, (h) Cutmix, (i) Ours+Cutmix, (j) MRS(label), (k) Ours+MRS(label).}
\label{MRBrainS13_Results}
\end{figure*}

\begin{figure}[t!]
    \centering
    \includegraphics[scale=0.48, trim=0cm 0.6cm 18.66cm 0cm, clip]{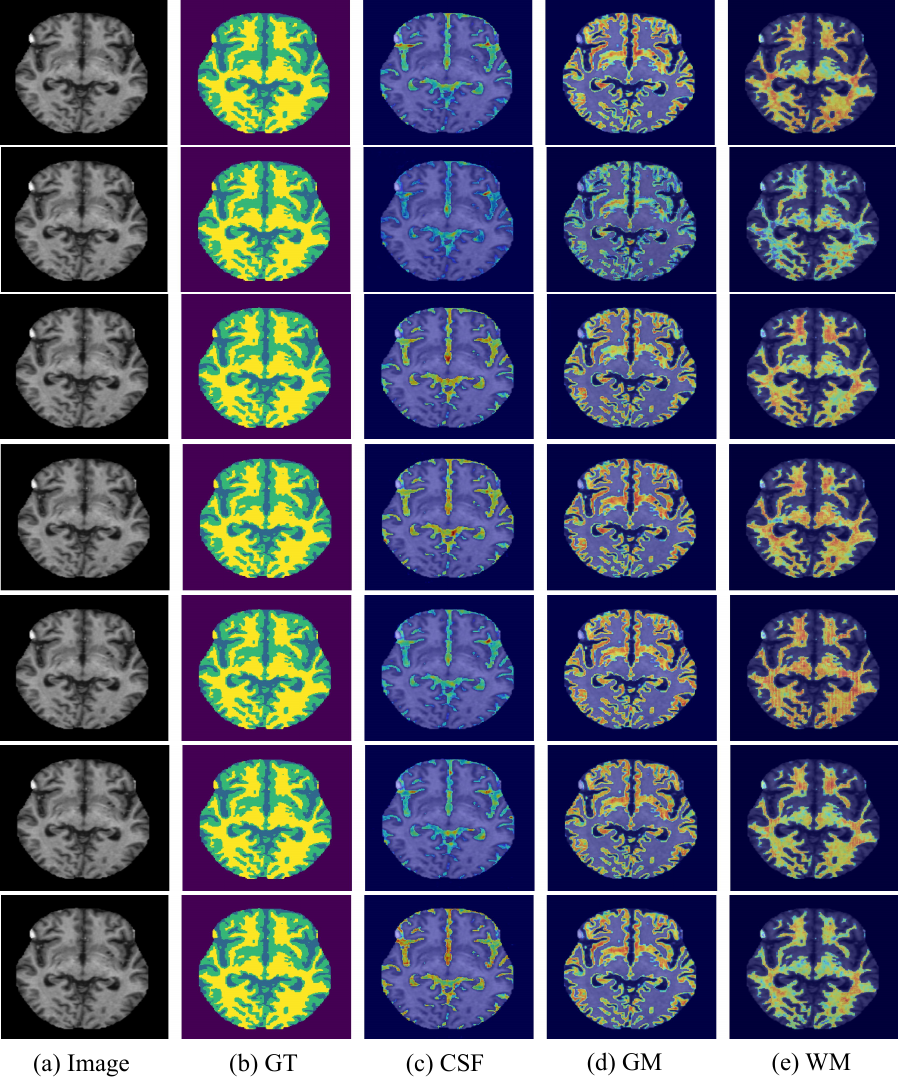}
    \caption{Class activation mapping (CAM) visualizations of different augmentation strategies on the OASIS-1 dataset. From top to bottom, the rows correspond to: Baseline, Random Erasing, Ours + Random Erasing, Cutout, Ours + Cutout, ABD (label), and Ours + ABD (label).}
    \label{fig:OASIS-1_cam}
\end{figure}

\begin{figure}[t!]
    \centering
    \includegraphics[scale=0.49, trim=0cm 1cm 17.35cm 0cm, clip]{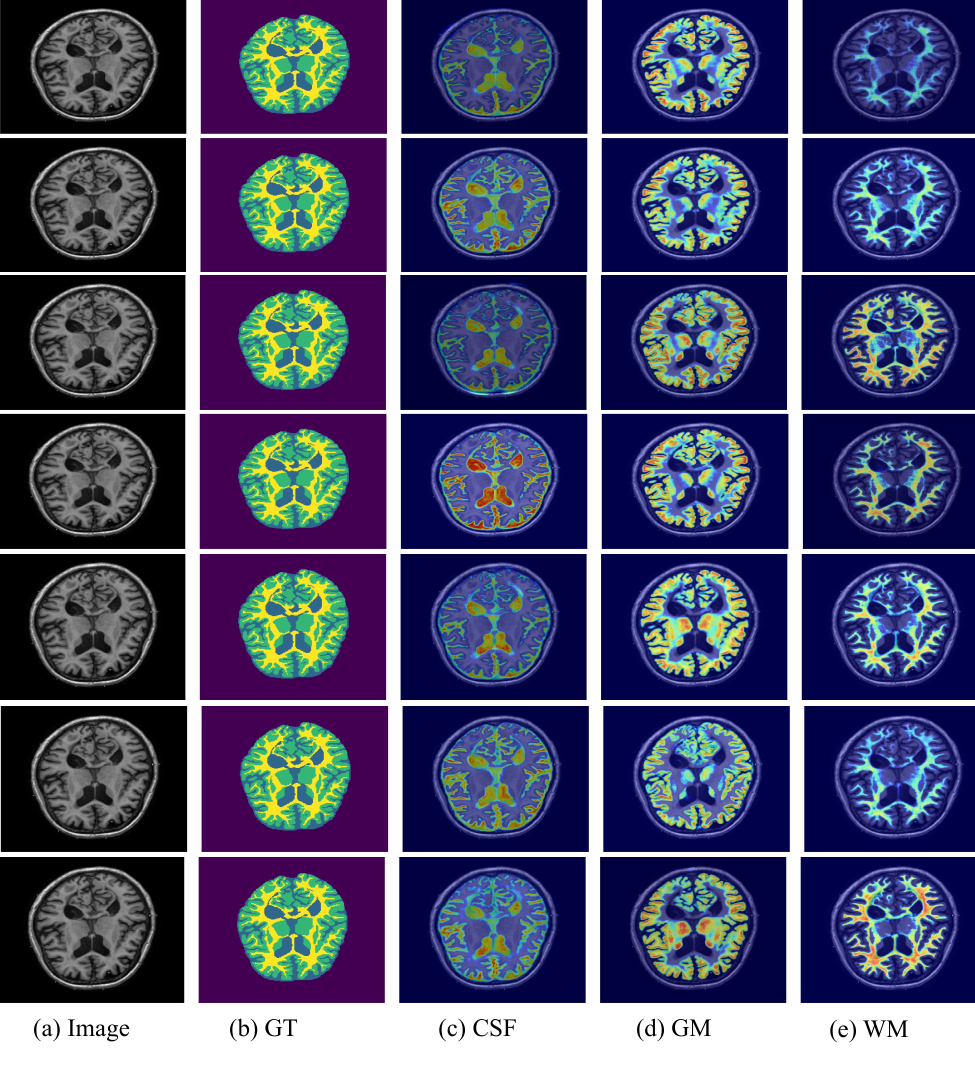}
    \caption{Class activation mapping (CAM) visualizations of different augmentation strategies on the MRBrainS13 dataset. From top to bottom, the rows correspond to: Baseline, Random Erasing, Ours + Random Erasing, Cutout, Ours + Cutout, ABD (label), and Ours + ABD (label).}
    \label{fig: MRBrainS13_cam}
\end{figure}

\begin{figure*}[t!]
    \centering
    \includegraphics[scale=0.65, trim=1cm 0cm 0.5cm 2.5cm, clip]{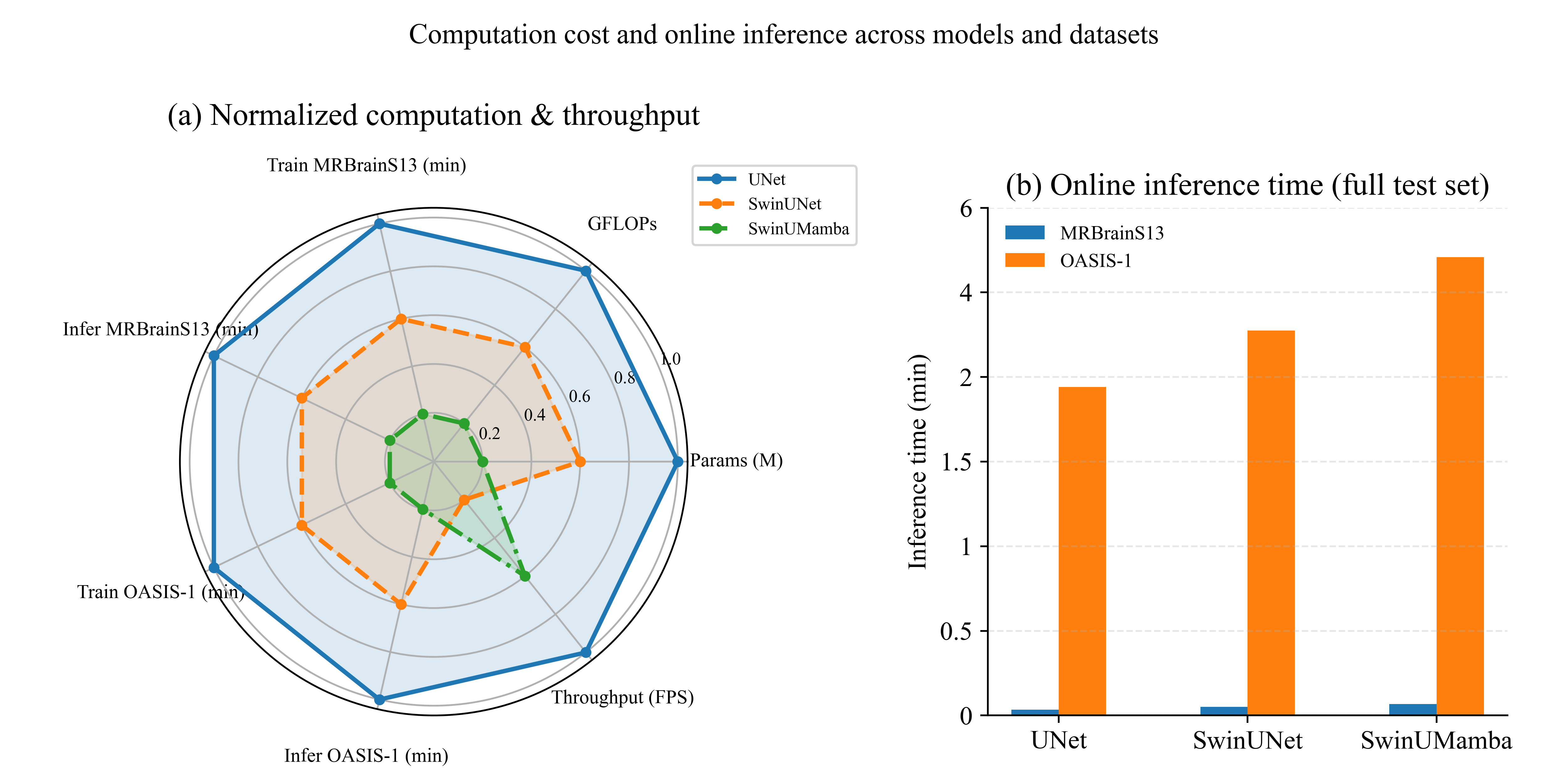}
    \caption{Computation cost across models, datasets, and online stages.
    (a) Normalized computation and throughput comparison across UNet, SwinUNet, and SwinUMamba, 
    considering model size, FLOPs, training times, inference times, and throughput.
    (b) Online inference time on the full test sets for MRBrainS13 and OASIS-1.}
    \label{fig:compute cost}
\end{figure*}


\end{document}